\pdfoutput=1
\documentclass[%
reprint,
superscriptaddress,
nofootinbib,
 amsmath,amssymb,
 aps,
 prx,
 longbibliography,
floatfix,
]{revtex4-1}

\usepackage{graphicx}
\usepackage{dcolumn}
\usepackage{bm}
\usepackage{appendix}
\usepackage{hyperref}
\usepackage{graphicx}
\usepackage{amsmath}
\usepackage{blkarray, bigstrut}
\usepackage{amsthm}
\usepackage{algorithm}
\usepackage{algpseudocode}

\usepackage{mathtools}

\usepackage{array,amssymb,booktabs}
\newcolumntype{C}{>{$}c<{$}}
\AtBeginDocument{
    \heavyrulewidth=.08em
    \lightrulewidth=.05em
    \cmidrulewidth=.03em
    \belowrulesep=.65ex
    \belowbottomsep=0pt
    \aboverulesep=.4ex
    \abovetopsep=0pt
    \cmidrulesep=\doublerulesep
    \cmidrulekern=.5em
    \defaultaddspace=.5em
}

\usepackage[usenames, dvipsnames]{color}

\newcommand\mybigstrut[1][4pt]{\setlength\bigstrutjot{#1}\bigstrut[t]}
\newcommand\mynegbigstrut[1][2pt]{\setlength\bigstrutjot{#1}\bigstrut[b]}

\begin{document}

\preprint{APS/123-QED}

\title{Parallel fault-tolerant programming of an arbitrary feedforward photonic network}

\author{Sunil Pai}
\email{sunilpai@stanford.edu}
\author{Ian A. D. Williamson}
\affiliation{Department of Electrical Engineering, Stanford University, Stanford, CA 94305, USA}
\author{Tyler~W.~Hughes}
\affiliation{Department of Applied Physics, Stanford University, Stanford, CA 94305, USA}
\author{Momchil~Minkov}
\author{Olav Solgaard}
\author{Shanhui Fan}
\author{David A. B. Miller}
\affiliation{Department of Electrical Engineering, Stanford University, Stanford, CA 94305, USA}
\begin{abstract}
    Reconfigurable photonic mesh networks of tunable beamsplitter nodes can linearly transform $N$-dimensional vectors representing input modal amplitudes of light for applications such as energy-efficient machine learning hardware, quantum information processing, and mode demultiplexing.
    Such photonic meshes are typically programmed and/or calibrated by tuning or characterizing each beam splitter one-by-one, which can be time-consuming and can limit scaling to larger meshes.
    Here we introduce a graph-topological approach that defines the general class of feedforward networks commonly used in such applications and identifies columns of non-interacting nodes that can be adjusted simultaneously.
    By virtue of this approach, we can calculate the necessary input vectors to program entire columns of nodes in parallel by simultaneously nullifying the power in one output of each node via optoelectronic feedback onto adjustable phase shifters or couplers. 
    This parallel nullification approach is fault-tolerant to fabrication errors, requiring no prior knowledge or calibration of the node parameters, and can reduce the programming time by a factor of order $N$ to being proportional to the optical depth (or number of node columns in the device).
    As a demonstration, we simulate our programming protocol on a feedforward optical neural network model trained to classify handwritten digit images from the MNIST dataset with up to 98\% validation accuracy.
\end{abstract}

\maketitle


\section{Introduction}

\begin{figure}[h]
    \centering
    \includegraphics[width=\linewidth]{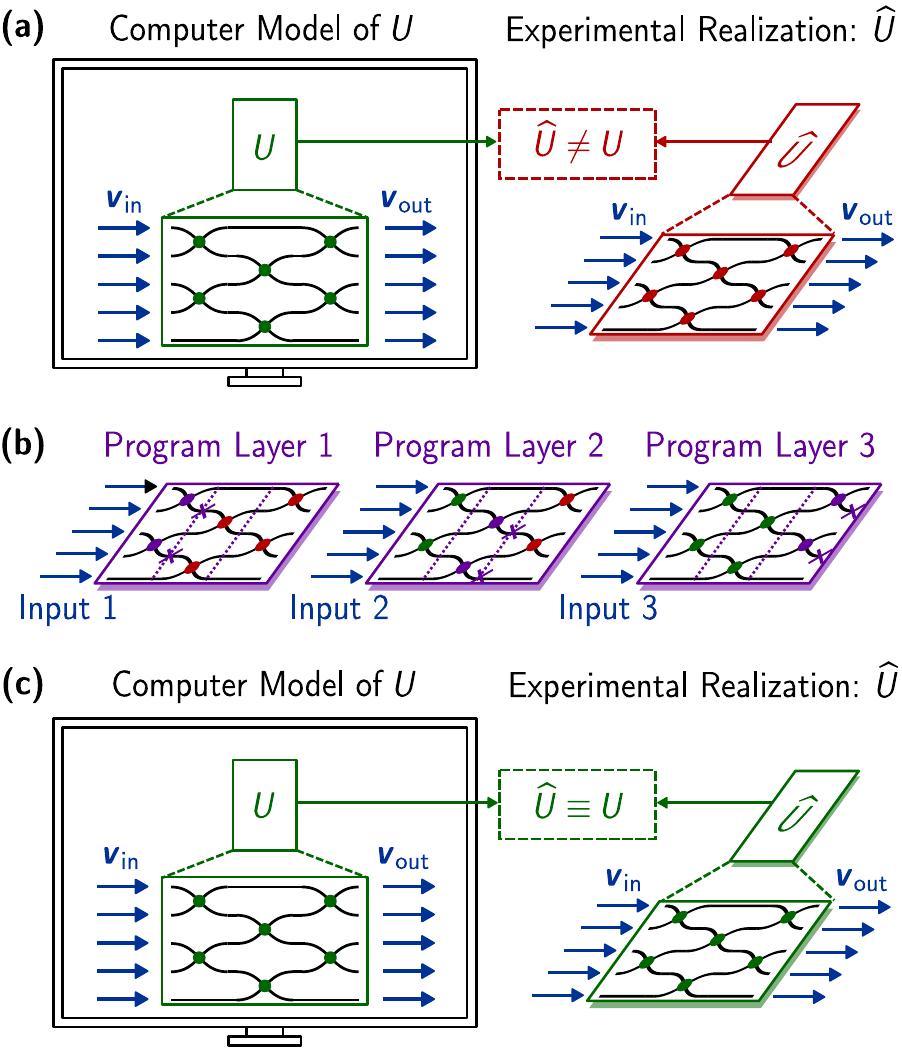}
    \caption{(a) Prior to programming the mesh to desired physical operator $\widehat{U}$ (shown as unprogrammed in red), we have our corresponding calculated mesh model $U$ (connectivity and tunable parameters) stored in the computer. (b) From $U$, we calculate a set of nullification vectors (the nullification set), and then shine in corresponding physical mode vectors from this set sequentially (Input 1, Input 2 and Input 3). We tune the beamsplitter parameters of the corresponding column (denoted by purple dots) by nullifying power at all sampling photodetectors (denoted by purple crosses) in the column in parallel. (c) After this programming, we are guaranteed that our experimental realization and computer model match up to an output phase reference, i.e. $\widehat{U} \equiv U$ (denoted as programmed in green).}
    \label{fig:motivation}
\end{figure}

Feedforward networks of tunable beamsplitters or ``nodes'', typically implemented as meshes of Mach-Zehnder interferometers (MZIs), can perform linear operations on sets or ``vectors'' of $N$ optical inputs in $N$ single-mode waveguides \cite{Reck1994ExperimentalOperator, Miller2013Self-configuringInvited, Carolan2015UniversalOptics}.
With advances in photonic integration, these networks have found a wide range of classical and quantum photonic applications where the mesh is configured to implement some specific linear transform or matrix. Some applications, like mode unscrambling in optical communications \cite{Annoni2017UnscramblingModes}, favor a self-configuring approach \cite{Miller2013Self-configuringInvited, Miller2013Self-aligningCoupler} in which the mesh sets itself up in real time to implement the matrix that undoes the mixing. Other applications, including photonic neural networks \cite{Shen2017DeepCircuits}, universal linear quantum computing \cite{Carolan2015UniversalOptics}, and photon random walks \cite{Harris2017QuantumProcessor}, may need to have the mesh implement some specific matrix that is calculated externally. These applications promise fast and energy-efficient matrix multiplication or analog computation via the physical process of light propagating through programmed nodes that can arbitrarily redistribute that light. For such applications, we will demonstrate how the nodes in an arbitrary feedforward network (e.g., the simple grid network of Fig. \ref{fig:motivation}) can be efficiently programmed in a fault-tolerant manner to implement a desired matrix operator.

Though it is straightforward to calculate phase shifts and/or beamsplitter split ratios to implement a matrix in such a mesh, any fixed fabrication of such settings is challenging for large meshes due to the precise settings required \cite{Flamini2017BenchmarkingProcessing}. For example, these errors limit the classification accuracy of optical neural networks \cite{Shen2017DeepCircuits, Fang2019DesignImprecisions} and prevent scaling up the number of components in quantum linear optical circuits \cite{Carolan2015UniversalOptics, Flamini2017BenchmarkingProcessing}. We therefore prefer reconfigurable beamsplitter nodes in such networks and corresponding setup algorithms that can directly program desired matrices.

Such setup algorithms also let us fully calibrate the nodes; finding the voltage drive settings for any programmed node subsequently allows us to interpolate among those settings to implement desired matrices by applying appropriate voltages directly. 
Although each node can be individually calibrated in some specific architectures \cite{Mower2015High-fidelityCircuits, Carolan2015UniversalOptics}, this approach is slow for large-scale meshes and must be repeated if components experience environmental drift. Here, we show an approach that can greatly reduce the time required for such setup and/or calibration processes and also generalizes to any feedforward mesh (i.e., where light propagates unidirectionally). By exploiting a graph-topological approach, we identify which nodes can be programmed simultaneously and provide the necessary parallelized algorithm to reduce the setup time by a factor of order $N$.

One general setup approach that does not require prior knowledge of each node's parameters is based on what could be called ``nullification.'' By specific choices of $N$ input modes (amplitude and phase) sent into $N$ input waveguides, corresponding nodes can be programmed by nullifying power at one of their outputs by optoelectronic feedback control \cite{Miller2013Self-configuringInvited, Annoni2017UnscramblingModes, Miller2017SettingMethod, Hughes2019ReconfigurableChip}.
For example, networks made from one or more diagonal sets of nodes \cite{Miller2013Self-configuringInvited} are programmed using nullification. One such diagonal gives a self-aligning beam coupler \cite{Miller2013Self-aligningCoupler}, and multiple diagonals can be cascaded to form triangular grid networks (as in the Reck scheme \cite{Reck1994ExperimentalOperator, Miller2013Self-configuringInvited, Annoni2017UnscramblingModes}) that implement arbitrary $N \times N$ unitary matrices. The appropriate input vectors programming such meshes are then the complex conjugates of the rows of the target unitary matrix \cite{Miller2013Self-configuringInvited, Annoni2017UnscramblingModes}.

More generally, other feedforward networks, such as rectangular grids (Clements scheme \cite{Clements2016AnInterferometers}), may not support self-configuration, so some separate design calculation must be performed to calculate the desired parameters of each node. Nonetheless, with the knowledge of the desired node parameters, such networks can be progressively configured using the reversed local light interference method (RELLIM) nullification approach proposed in Ref. \citenum{Miller2017SettingMethod}. A key question is whether we can minimize the total time required to program or calibrate the network. The self-configuring algorithm for diagonal or triangular meshes and the RELLIM algorithm as originally conceived \cite{Miller2017SettingMethod} give prescriptions for setting up the nodes sequentially (i.e., one-at-a-time) and thus require a number of steps equal to the number of nodes.

In this paper, we propose a graph-topological framework that arranges any given feedforward network into columns of nodes that can be programmed simultaneously, with just one ``nullification set'' input vector for the entire column, rather than programming one node at a time with possibly different input vectors for each such node. The resulting ``parallel nullification'' or parallel RELLIM (``PRELLIM'') protocol uses up to ($N / 2$)-times fewer calibration steps and input vectors than RELLIM, where $N$ is the number of input modes to the system. Our protocol ultimately enables efficient, fault-tolerant, flexible, and scalable calibration (with time complexity of the number of columns in the device) of an \textit{arbitrary} feedforward photonic mesh architecture. Example such architectures include triangular \cite{Reck1994ExperimentalOperator, Miller2013Self-configuringInvited} and rectangular \cite{Clements2016AnInterferometers} grid networks (capable of implementing arbitrary unitary matrices) and butterfly \cite{Flamini2017BenchmarkingProcessing, Fang2019DesignImprecisions} networks (capable of implementing any permutation or DFT unitary matrices).

We outline a typical scenario that benefits from parallel nullification in Fig. \ref{fig:motivation}(a), where a model of an optical network stored in a digital computer (e.g. a CPU-trained optical neural network) must be programmed into a photonic circuit.
The model consists of the network topology (node connection patterns) and tunable node parameters used to calculate the nullification set vectors for calibration. 
The parallel nullification procedure shown in Fig. \ref{fig:motivation}(b), consists of calibrating the mesh one column at a time using the nullification set and tuning all nodes within each column in parallel until their bottom outputs are all nullified (i.e. transmit zero power). 
In Fig. \ref{fig:motivation}(c), we show that after parallel nullification is applied to all columns, our device matches the computer model as accurately as physically possible.

The remainder of this paper is organized as follows. In Section \ref{sec:photonicmesh}, we lay out the foundations of our graph-topological framework used to formally define a general feedforward photonic mesh. In Section \ref{sec:parallelnullification}, we propose our parallel nullification protocol and demonstrate how our protocol can deploy machine learning models on optical neural networks in Section \ref{sec:apps}. We then more generally discuss fault-tolerant performance of parallel nullification in presence of systematic errors in Section \ref{sec:error}.

\section{Feedforward photonic network}
\label{sec:photonicmesh}

In this section, we introduce some required mathematical and graph-topological terminology and concepts for feedforward photonic networks. For any such network with $N$ input and output waveguides, there is a linear device operator \cite{Miller2012AllConverters, Miller2019WavesOptics} or matrix $U$ relating the input and output waveguide amplitudes for monochromatic light at steady state. Because we are considering feedforward networks, by choice we consider only the forward amplitudes in all waveguides. With a set of $N$ amplitudes in the input waveguides, represented by the input mode vector $\boldsymbol{v}_{\mathrm{in}} \in \mathbb{C}^N$ (the $n$th element stores the amplitude and phase of the input mode in waveguide $n$), and a corresponding vector of output mode amplitudes $\boldsymbol{v}_{\mathrm{out}}$, then $\boldsymbol{v}_{\mathrm{out}} = U \boldsymbol{v}_{\mathrm{in}}$ as shown in Fig. \ref{fig:motivation}. Since we presume no backwards waves, $U$ can be considered to be an $N \times N$ transmission matrix. 
Each individual tunable beamsplitter node (depicted by dots in Fig. \ref{fig:motivation}) can similarly be described by a $2 \times 2$ transmission matrix $T_2$ that describes how the light in its two input waveguide ports is distributed across its two output ports.  

Based on the graph-topological arguments of this section, we always arrive at a compact definition for $U$ in terms of node columns that can each be programmed in a single step as suggested by Fig. \ref{fig:motivation}. For example, the triangular grid mesh \cite{Reck1994ExperimentalOperator} can be programmed in $2N$ steps (as opposed to the typical $N(N-1)/2$ steps), the rectangular grid mesh \cite{Clements2016AnInterferometers} in $N$ steps, and the butterfly mesh \cite{Flamini2017BenchmarkingProcessing} in $\log{N}$ steps.

\subsection{Nodes}

Each node of the feedforward network is a $2 \times 2$ tunable beamsplitter whose requirement is to be able to arbitrarily redistribute light. This is accomplished by concatenating a phase shifter ($S_{\mathrm{P}}(\phi)$) to a tunable coupler ($S_{\mathrm{A}}(\theta)$) resulting in the transmission matrix $T_2$:
\begin{equation} \label{eqn:tunablebeamsplitter}
    \begin{aligned}
    T_2(\theta, \phi) &= S_{\mathrm{A}}(\theta) S_{\mathrm{P}}(\phi) \equiv i\begin{bmatrix}e^{i\phi}\sin \frac{\theta}{2} & \cos \frac{\theta}{2} \\ e^{i\phi}\cos \frac{\theta}{2} & -\sin \frac{\theta}{2}
    \end{bmatrix},
    \end{aligned}
\end{equation}
where $\theta \in [0, \pi]$ and $\phi \in [0, 2\pi)$. By notation ``$\equiv$,'' we note that there are many equivalent constructions of $T_2$ provided explicitly in Appendix \ref{sec:tb}, which have mathematically different, but functionally equivalent, representations including tunable directional couplers (TDCs) and Mach-Zehnder interferometers (MZIs).

Parallel nullification is agnostic of the exact modulation schemes for $S_\mathrm{P}$ and $S_\mathrm{A}$, as long as $S_\mathrm{P}$ corresponds to a controllable phase difference between the two input waveguides (so a differential phase) and $S_\mathrm{A}$ covers the full range of transmissivities (bar state or $\theta = \pi$ to cross state or $\theta = 0$).

In an $N$-port device, to represent the effect of one $2 \times 2$ element we may embed the $2 \times 2$ transmission matrix $T_2$ along the diagonal of an otherwise $N \times N$ identity matrix. Formally, this would allow the resulting embedded operation $T_N^{[m]}$ to operate between modes in waveguides $2m - 1$ and $2m$ (a Givens rotation) as follows:


\begin{equation}\label{eqn:givensrotation}
T_N^{[m]} \,:=\,\, \begin{blockarray}{cccccccc}
 &  & \small{2m - 1} & \small{2m} &  & & \\
\begin{block}{[cccccc]cl}
1   & \cdots &    0   &   0   & \cdots &    0 & & \mybigstrut\\
\vdots & \ddots & \vdots &  \vdots &        & \vdots & &\\
0   & \cdots &    T_{11}  &   T_{12}   & \cdots &    0 & & \small{2m - 1} \\
0   & \cdots &    T_{21}  &   T_{22}  & \cdots &    0   & & \small{2m}\\
\vdots &   & \vdots  & \vdots & \ddots & \vdots & &\\
0   & \cdots &   0 &    0   & \cdots &    1 & & \mynegbigstrut\\
\end{block}
 &  & & &  & & \\
\end{blockarray}.\\
\end{equation}

\begin{figure}[h]
    \centering
    \includegraphics[width=\linewidth]{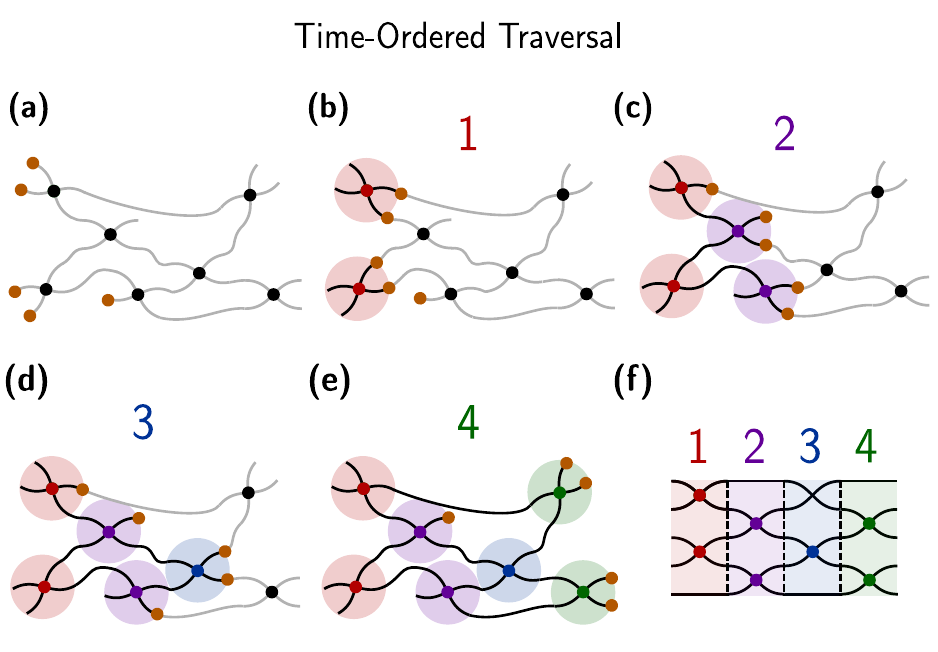}
    \caption{From ``left'' to ``right,'' we mathematically propagate a 5-mode vector of amplitudes (time-ordered waveguide traversal of orange dots) on an arbitrary $N = 5$ feedforward mesh (a) over panels (b) - (e) to generate the final compact, ``layered'' configuration of this mesh in panel (f). The numbers indicate the time step in which the nodes (shaded by a colored circle) are being traversed (i.e., simultaneously configured), and these numbers ultimately correspond to the columns (or time steps) to which nodes traversed at that step belong. The rule in the traversal is that two orange dots need to be at the input of a given node before that node can be traversed (i.e., the orange dots advance past the node) for that time step. (Note at the top of column 3 we have two waveguides crossing; there is no node at this crossing, and the waveguides cross without any interaction.)}
    \label{fig:meshtopology}
\end{figure}

\subsection{Graph-topological framework}

We typically make photonic mesh networks in a grid-like manner \cite{Reck1994ExperimentalOperator, Miller2013Self-configuringInvited, Clements2016AnInterferometers} for efficiency and compactness in fabrication. This also can allow equal path lengths if we want to make the interference relatively insensitive to wavelength changes. Configuring the nodes of the network takes much longer than the time for light to propagate through the network. So, for the purposes of analyzing more general feedforward networks, we can consider monochromatic, continuous-wave light such that actual optical distances and geometrical arrangements are unimportant for calibration and programming of the network. 

In contrast, the ``network topology'' (i.e., connectivity independent of lengths of inter-node links) \textit{is} important for defining valid feedforward configurations and the required inputs and node sequence by which we can progressively program or calibrate the network. We could mathematically view the inter-node links of the network as made of flexible (and stretchable) fibers between nodes that can move in space as long as the network topology is not changed. As such, we represent the notion of light moving ``forward'' or left-to-right (independent of the physical node locations in space) as moving along a given fiber away from network inputs or node outputs, which we define to be on the ``left,'' and towards network outputs or node inputs, which we define to be on the ``right.'' We will ultimately examine grouping nodes into ``columns'' to establish which nodes can be configured simultaneously or within the same propagation time step.

In such a graph-topological view, we can propose a sequential constructive definition that generates any arbitrary feedforward mesh network. As represented in the example in Fig. \ref{fig:meshtopology}(a), we start with $N = 5$ input waveguides or fibers on the physical left (though this positioning is arbitrary). In constructing the network, we perform node addition by interfering any chosen pair of waveguides or fibers at the inputs of a node to generate outputs from that node. We assume light flows forward, so at any subsequent (further to the right) node addition, we interfere any two waveguides that exit earlier (further to the left) nodes. In this way, given a sufficient number of nodes and arbitrary choice of waveguides to couple, we can generate any feedforward mesh architecture. Furthermore, rather than adding a single node at a time in our construction, we can add at most $M := \lfloor N / 2 \rfloor$ nodes at a time (where $\lfloor x \rfloor$ represents the largest integer less than or equal to $x$) to account for the largest number of waveguides that can be simultaneously coupled. The resulting architecture is shown in Fig. \ref{fig:meshtopology}(f) where we maintain the feedforward property that propagation along the $N = 5$ waveguides always crosses the dotted vertical lines from left-to-right. This results in what we will define as the most ``compact'' representation, resembling more closely some of the commonly known rectangular and triangular feedforward architectures \cite{Reck1994ExperimentalOperator, Clements2016AnInterferometers}.

It can be helpful to find this compact representation of Fig. \ref{fig:meshtopology}(f) starting from the more arbitrary layout of Fig. \ref{fig:meshtopology}(a) using a ``topological sort'' or homomorphic transformation. To do this, we propose a time-ordered ``traversal'' where we hypothetically insert a mode at every input to the mesh and allow the resulting mode vector mathematically to propagate progressively (i.e., to traverse) through the device. At every step of the traversal, shown in Fig. \ref{fig:meshtopology}(a)-(e), we allow the mode vector amplitudes to pass through and be transformed by the nodes in the network. Since the transmission matrix for each node in Eq. \ref{eqn:tunablebeamsplitter} interferes two inputs, two modes need to be input into a node to traverse that node (advance to the outputs of that node). Via this procedure, we equivalently construct sets of nodes (in circles of the same color of Fig. \ref{fig:meshtopology}) that all connect only to previous (already traversed) nodes. We can then ``compactify'' the network into the numbered vertical columns as in Fig. \ref{fig:meshtopology}(f) corresponding to the time steps (colored and indexed 1 to 4) in which nodes are traversed. Since the nodes in these columns are not connected to one another, we are free to configure these nodes in any order, including configuring them all simultaneously (i.e., in the same time step) assuming the preceding nodes are configured. Using this graph-topological approach, we can always generate the most compact device (in terms of number of node columns or ``optical depth'') possible for any feedforward architecture by lining up the nodes according to their time step as in Fig. \ref{fig:meshtopology}(f). 

Such a compactified version is the network with the lowest optical depth representation of the mesh, and nodes in each time step (or column) must be independent since there cannot be a valid path between them without contradicting our traversal algorithm; hence they can to be tuned in parallel. Our column-wise or timestep labelling is a specific case of the well-known Dijkstra's algorithm \cite{Dijkstra1959AGraphs} to find the maximum longest path (over all input source nodes) in a directed acyclic graph (feedforward mesh) using breadth-first (time-order) search to the mesh.

Each node belongs to exactly one column, but if our traversal algorithm finds multiple column assignments for that node (which would require revisiting that node), then there would be be a cycle in the graph. Therefore, by applying this algorithm, we can formally identify cycles in any mesh architecture (as in the lattice meshes of Refs. \citenum{Perez2017MultipurposeCore, Perez2019ScalableMeshes}) that would disqualify such a mesh as a feedforward architecture. Though those meshes with cycles have specific uses, nodes in such meshes must be tuned individually or using some global optimization approach \cite{Perez2019ScalableMeshes} and may be susceptible to back-reflections during programming or calibration.

\subsection{Transmission matrix representation}

\begin{figure}[h]
    \centering
    \includegraphics[width=\linewidth]{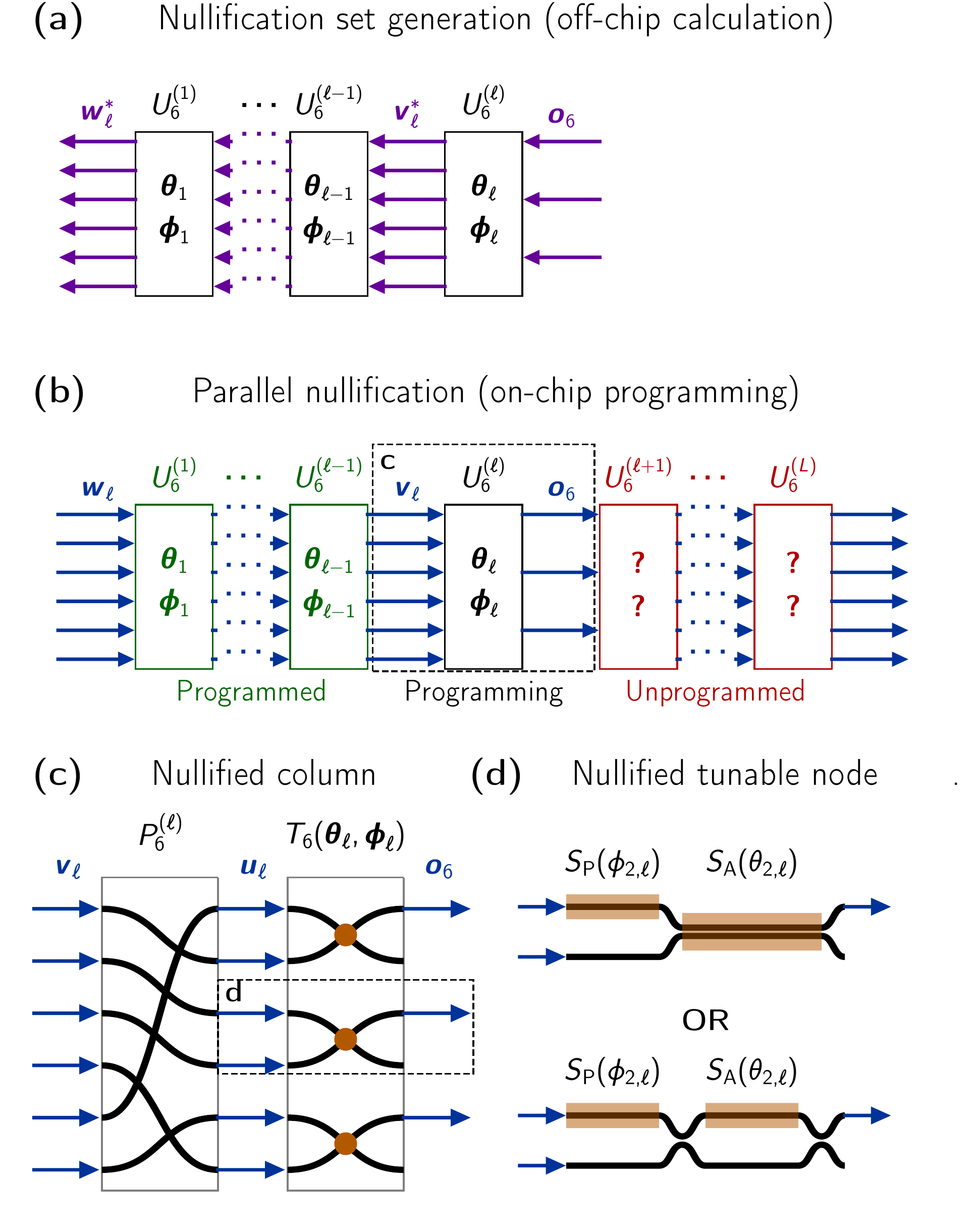}
    \caption{We present the parallel nullification protocol for any feedforward photonic mesh network, here for an example with $N = 6$ waveguides. (a) Nullification set generation for column $\ell$ in a mesh using RELLIM \cite{Miller2017SettingMethod}. (b) Parallel nullification of column $\ell$ using nullification set vector $\boldsymbol{w}_\ell$ assumes columns $1, 2, \ldots \ell - 1$ have already been programmed. (c) Parallel nullification at column $\ell$ tunes $\boldsymbol{\theta}_\ell, \boldsymbol{\phi}_\ell$ in parallel via independent optoelectronic feedback optimizations until all bottom ports are nullified. (d) Nullification requires phase equalization ($S_P$) and split ratio modulation ($S_A$).}
    \label{fig:parallelnullification}
\end{figure}

Although our graph-topological construction can generally be used to define any feedforward mesh, we need an equivalent transmission matrix that allows for straightforward simulation and calibration of such devices. Once we have arrived at the compact representation of Fig. \ref{fig:meshtopology}(f), we can define each column entirely using the general definition of Eq. \ref{eqn:ffmeshcolumn}, later depicted in Fig. \ref{fig:parallelnullification}(c). For each added column of nodes, we select the appropriate source nodes using the permutation matrix $P_N^{(\ell)}$ and connect them to simultaneously-acting nodes via the block-diagonal matrix $T_N(\boldsymbol{\theta}_\ell, \boldsymbol{\phi}_\ell)$ (the product of up to $M = \lfloor N / 2 \rfloor$ nodes given by Eq. \ref{eqn:givensrotation}):
\begin{equation} \label{eqn:ffmeshcolumn}
    \begin{aligned}
        T_N(\boldsymbol{\theta}_\ell, \boldsymbol{\phi}_\ell) &:= \prod_{m = 1}^{M} T_{N}^{[m]}(\theta_{m, \ell}, \phi_{m, \ell})\\
        U_N^{(\ell)} &:= T_N(\boldsymbol{\theta}_\ell, \boldsymbol{\phi}_\ell) P_N^{(\ell)},
    \end{aligned}
\end{equation}
where phase parameters are $\boldsymbol{\theta}_\ell = (\theta_{1, \ell}, \ldots \theta_{m, \ell}, \ldots \theta_{M, \ell})$ and $\boldsymbol{\phi}_\ell = (\phi_{1, \ell}, \ldots \phi_{m, \ell}, \ldots \phi_{M, \ell})$. Assuming $M_\ell \leq M$ nodes in the column, then $P_N^{(\ell)}$ is defined such that we can add synthetic bar state beamsplitter nodes for all $m > M_\ell$, i.e. $\theta_{m, \ell} = \phi_{m, \ell} = \pi$ for any waveguides that do not interact in that column. While all waveguides in a column can technically be interfered at nodes for even $N$, for odd $N$ there will always be a single remaining waveguide that is not connected to a node in that column. However, this ultimately does not change the definition in Eq. \ref{eqn:ffmeshcolumn} since we can always choose this to be the $N$th waveguide by appropriate choice of $P_N^{(\ell)}$. 

To complete meshes for which we desire a fully arbitrary unitary transformation, we may need a further set of phase shifters on the output nodes of the mesh that set the relative phases of the rows of the implemented matrix (represented by the diagonal unitary matrix $\Gamma_N(\boldsymbol{\gamma})$). (Such phase shifters can also be at the input if the external phase shift of each node is applied \textit{after} the tunable coupler, effectively a mirror image of our current definition.) Furthermore, we include a final permutation $P_N$ before or after those final phase shifters, allowing us to arbitrarily rearrange the rows of the matrix at the end.

Performing our transmission matrix construction for $L$ columns, we arrive at an expression for an arbitrary feedforward mesh:
\begin{equation}\label{eqn:ffmesh}
    \begin{aligned}
        U_N(\boldsymbol{\theta}, \boldsymbol{\phi}, \boldsymbol{\gamma}) &:= D_N \prod_{\ell=1}^{L} U_N^{(\ell)}\\
        D_N &:= \Gamma_N(\boldsymbol{\gamma})P_N,
    \end{aligned}
\end{equation}
where the matrices $\boldsymbol{\theta} := \{\theta_{m, \ell}\}, \boldsymbol{\phi} := \{\phi_{m, \ell}\}$ and vector $\boldsymbol{\gamma} := \{\gamma_{n}\}$ represent the full set of mesh parameters with ranges $\theta_{m, \ell} \in [0, \pi]$ and $\phi_{m, \ell}, \gamma_{n} \in [0, 2\pi)$. Any feedforward architecture is entirely defined by these permutation matrices $\{P_N^{(\ell)}\}$ and $P_N$ (representing the choices of waveguides to interfere), which we explicitly define for commonly proposed architectures (e.g., triangular, rectangular, butterfly) in Appendix \ref{sec:meshexamples}, and by the various phase settings in the nodes and at the output.

\section{Parallel nullification} \label{sec:parallelnullification}

Now that we have defined a column-wise feedforward photonic mesh, we present parallel nullification as summarized in Fig. \ref{fig:parallelnullification}. The programming algorithm consists of an off-chip calculation of a sequence of inputs (or ``nullification set'') as in Fig. \ref{fig:parallelnullification}(a) followed by parallel nullification of columns from left-to-right of the mesh network as in Fig. \ref{fig:parallelnullification}(b). 

In a realistic setting, we do not have direct access to the input of each column, but rather the inputs to the overall device. We therefore need the overall input vector assigned to each column $\ell$ that leads to the desired nullified output for that column. In both the RELLIM protocol \cite{Miller2017SettingMethod} and our parallel nullification approach introduced here, the nullification set calculation works due to the reciprocity of feedforward meshes. In particular, we calculate a vector of complex amplitudes that would emerge from shining the desired nullified output backwards to the input from any column $\ell$ through a correctly programmed mesh. The nullification set for RELLIM results from mathematically propagating light backwards from the desired output of a \textit{single} nullified node \cite{Miller2017SettingMethod}. In contrast, the nullification set for parallel nullification results from mathematically propagating light backwards from the desired output of an entire \textit{column} of nullified nodes, as in Fig. \ref{fig:parallelnullification}(a). 

When it is time to actually program the columns on the physical mesh, we physically send the phase-conjugate (i.e., complex conjugate) of this result, which we can now call the nullification set vector $\boldsymbol{w}_\ell$, into the mesh inputs as in RELLIM 
\cite{Miller2017SettingMethod}. As long as all the preceding mesh columns are set correctly, reciprocity ensures this vector can be used to correctly program the corresponding column $\ell$ to the desired mesh settings as shown in Fig. \ref{fig:parallelnullification}(b)-(d) by physically nullifying the appropriate outputs of that column.

In this section, we first formalize the nullification set calculation which is performed separately on a traditional computer. We then discuss the mathematics and physical procedures behind parallel nullification of each column and the overall programming algorithm that sets the physical parameters of the device (which we will refer to as $\boldsymbol{\alpha}, \boldsymbol{\beta}$) to the desired settings ($\boldsymbol{\theta}, \boldsymbol{\phi}$ respectively).

\subsection{Nullification set calculation}

For parallel nullification, there exist many valid calculations of nullification sets; we could choose to nullify either the top or the bottom port at any given node, and the value of the (non-zero) power at the non-nullified port does not matter. For definiteness, we will consider a simple valid target vector $\boldsymbol{o}_N = (1, 0, 1, 0, \cdots)$ or more formally:
\begin{equation}
\boldsymbol{o}_N := \sum_{m = 1}^{M} \boldsymbol{e}_{2m - 1},
\end{equation}
where $\boldsymbol{e}_{2m - 1}$ represents the $(2m - 1)$th standard Euclidean basis vector in $\mathbb{C}^N$, or equivalently, unit power in node output port $2m - 1$.

We calculate the nullification set vector $\boldsymbol{w}_\ell$ for each column $\ell$ as depicted in Fig. \ref{fig:parallelnullification}(a):
\begin{equation} \label{eqn:nullvector}
\boldsymbol{w}_\ell^* := \prod_{\ell' = 1}^\ell \left(U_N^{(\ell')}\right)^T \boldsymbol{o}_N.
\end{equation}
Since we have already programmed columns $1 \to \ell - 1$, sending in $\boldsymbol{w}_\ell$ to the device yields correct values for $\boldsymbol{\theta}_\ell, \boldsymbol{\phi}_\ell$ after parallel nullification of column $\ell$.

We can compute the entire nullification set $\{\boldsymbol{w}_1, \boldsymbol{w}_2, \ldots \boldsymbol{w}_L\}$ off-chip in $O(N \cdot L^2)$ time assuming all $\boldsymbol{\theta}, \boldsymbol{\phi}$ are known since each $\boldsymbol{w}_\ell$ results from reverse propagating $\boldsymbol{o}_N$ through $\ell$ columns. For example, we can calculate the nullification set in $O(N^3)$ time for the rectangular or triangular grid meshes. Code for calculating the nullification set (Eq. \ref{eqn:nullvector}) for any feedforward mesh is provided in our Python software module \href{https://github.com/solgaardlab/neurophox}{\texttt{neurophox}} \cite{Pai2019MatrixDevices}, further discussed in Appendix \ref{sec:neurophox},\footnote{See \href{https://github.com/solgaardlab/neurophox}{https://github.com/solgaardlab/neurophox}} and nullification set results calculated for rectangular grid networks \cite{Clements2016AnInterferometers} are shown in Appendix \ref{sec:nullset}.
\subsection{Nullification}

After calculating our nullification set off-chip, we program the physical device. Before programming a given node $m$ in column $\ell$, the settings $\alpha_{m, \ell}$ and $\beta_{m, \ell}$ will be different from desired settings $\theta_{m, \ell}$ and $\phi_{m, \ell}$ respectively. Given nullification set input $\boldsymbol{w}_\ell$ (presuming all preceding columns of the mesh are already set correctly), nullification can be achieved by two independent steps to nullify the bottom port (port $2m$) \cite{Miller2013Self-configuringInvited, Miller2013Self-aligningCoupler} as proven explicitly in Appendix \ref{sec:nullificationproof}:
\begin{enumerate}
    \item Sweep $\beta_{m, \ell}$ (i.e., adjust the relative phase of the node inputs) until bottom port power is \textit{minimized}.
    \item Sweep $\alpha_{m, \ell}$ (i.e., adjust the node split ratio) until bottom port is \textit{nullified}, as shown in Fig. \ref{fig:parallelnullification}(d).
\end{enumerate}
Given the permuted mode pair entering from the previous column $u_{2m - 1, \ell}, u_{2m, \ell}$, this straightforward two-step optimization exactly adjusts settings $\alpha_{m, \ell}$, $\beta_{m, \ell}$ to be the desired $\theta_{m, \ell}, \phi_{m, \ell}$:
\begin{equation} \label{eqn:nullification}
    \begin{aligned}
    \alpha_{m, \ell}^{\mathrm{opt}} &= 2 \arctan \left|\frac{u_{2m - 1, \ell}}{u_{2m, \ell}}\right| = \theta_{m, \ell}\\
    \beta_{m, \ell}^{\mathrm{opt}} &= -\mathrm{arg}\left(\frac{u_{2m - 1, \ell}}{u_{2m, \ell}}\right) = \phi_{m, \ell}.
    \end{aligned}
\end{equation}

Parallel nullification of all nodes in column $\ell$ (i.e. $T_N(\boldsymbol{\theta}_\ell, \boldsymbol{\phi}_\ell)$) can be achieved because, as discussed in Section \ref{sec:photonicmesh}, the choice of nodes assigned to column $\ell$ ensures all such optimizations are independent (i.e., do not influence each other). Nullification can be accomplished physically by sampling and measuring a small fraction of the power in the bottom output port. Nullification is then achieved through local feedback loops and is accomplished once zero power is measured. This nullification procedure has been experimentally demonstrated previously with noninvasive CLIPP detectors \cite{Annoni2017UnscramblingModes, Grillanda2014Non-invasiveElectronics, Morichetti2014Non-InvasiveMonitoring} that are effectively low-loss because they rely on light already absorbed in background loss processes in the waveguide.

\subsection{Programming algorithm}

\begin{algorithm}[H]
    \caption{Parallel nullification}
    \label{alg:parallelnullification}
    \begin{algorithmic}[1]
        \Function{NullificationVector}{$\boldsymbol{\theta}$, $\boldsymbol{\phi}$, $\ell$}
            \State $\boldsymbol{w}_{\mathrm{\ell}}^* \gets \boldsymbol{o}_N$
            \For{$\ell' \in [1, 2, \ldots, \ell]$}
                \State $\boldsymbol{w}_{\mathrm{\ell}}^* \gets \left(U_N^{(\ell - \ell')}\right)^T \boldsymbol{w}_{\mathrm{\ell}}^*$ \Comment Eq. \ref{eqn:nullvector}
            \EndFor
            \State \Return $\boldsymbol{w}_{\mathrm{\ell}}$
        \EndFunction
        \item[]
        \Function{ForwardPropagate}{$\boldsymbol{w}_\ell$, $\boldsymbol{\alpha}$, $\boldsymbol{\beta}$, $\ell$}
            \State $\boldsymbol{v}_{\ell} \gets \boldsymbol{w}_\ell$ 
            \For{$\ell' \in [1, \ldots, \ell - 1]$}
                \State $\boldsymbol{v}_{\ell} \gets \widehat{U}_N^{(\ell')} \boldsymbol{v}_{\ell}$ \Comment Eq. \ref{eqn:vectonull}
            \EndFor
            \State \Return $\boldsymbol{v}_{\ell}$
        \EndFunction
        \item[]
        \Procedure{ParallelNullification}{$\boldsymbol{\theta}$, $\boldsymbol{\phi}$}
            \For{$\ell \in [1, 2, \ldots, L]$} \Comment{Off-chip}
                \State $\boldsymbol{w}_{\ell} \gets $ \textsc{NullificationVector}($\boldsymbol{\theta}$, $\boldsymbol{\phi}$, $\ell$)
            \EndFor
            \For{$\ell \in [1, 2, \ldots, L]$} \Comment{On-chip}
                \State $\boldsymbol{v}_{\ell} \gets $ \textsc{ForwardPropagate}($\boldsymbol{w}_{\ell}$, $\boldsymbol{\alpha}$, $\boldsymbol{\beta}$, $\ell$)
                \State $\boldsymbol{u}_\ell \gets P_N^{(\ell)} \boldsymbol{v}_\ell$
                \For{$m \in [1, 2, \ldots, M_\ell]$} \Comment{In parallel}
                 \State $\alpha_{m, \ell} \gets \theta_{m, \ell}$
                 \State $\beta_{m, \ell} \gets \phi_{m, \ell}$ \Comment Eq. \ref{eqn:nullification}
                \EndFor
            \EndFor
        \EndProcedure
    \end{algorithmic}
\end{algorithm}

The parallel nullification programming algorithm proceeds formally as in Alg. \ref{alg:parallelnullification}, which we simulate in \href{https://github.com/solgaardlab/neurophox}{\texttt{neurophox}} \cite{Pai2019MatrixDevices}. If we are configuring the actual physical network, the entire \textsc{ForwardPropagate} method is a \textit{physical process} in which we generate an actual vector of optical inputs $\boldsymbol{w}_\ell$, and propagate them through the mesh to physically generate the vector $\boldsymbol{v}_{\ell}$ at the node outputs in layer $\ell - 1$, which are permuted by $P_N^{(\ell)}$ to produce the vector we nullify in Eq. \ref{eqn:nullification}, $\boldsymbol{u}_{\ell}$. The vector $\boldsymbol{v}_{\ell}$ is the propagated fields of our calculated inputs $\boldsymbol{w}_\ell$ to the output of column $\ell - 1$ as depicted in Fig. \ref{fig:parallelnullification}(b):
\begin{equation}\label{eqn:vectonull}
    \boldsymbol{v}_\ell := \prod_{\ell' = 1}^{\ell - 1} \widehat{U}_N^{(\ell - \ell')} \boldsymbol{w}_\ell,
\end{equation}
where we use $\widehat{U}_N^{(\ell)}$ to mathematically represent the transmission matrices describing already-programmed physical columns of nodes (correctly set to column parameters $\boldsymbol{\theta}_\ell, \boldsymbol{\phi}_\ell$). The final (parallel) for-loop of Alg. \ref{alg:parallelnullification} represents a physical parallel nullification of output powers using optoelectronic feedback on each column in order from $\ell = 1$ to $L$.

The inputs to Alg. \ref{alg:parallelnullification} are the desired settings for the feedforward mesh architecture which are used to first generate the nullification set $\{\boldsymbol{w}_1, \boldsymbol{w}_2, \ldots \boldsymbol{w}_L\}$. Algorithm \ref{alg:parallelnullification} ultimately results in setting the physical mesh parameters $\boldsymbol{\alpha}, \boldsymbol{\beta}$ to the desired $\boldsymbol{\theta}, \boldsymbol{\phi}$, for which a full testing suite is provided in in \href{https://github.com/solgaardlab/neurophox}{\texttt{neurophox}} \cite{Pai2019MatrixDevices}. As shown in Fig. \ref{fig:parallelnullification}(a), each nullification set vector $\boldsymbol{w}_\ell$ has the necessary information from past columns to tune all devices in column $\ell$ in parallel. We demonstrate the parallel nullification algorithm from Fig. \ref{fig:parallelnullification}(b) for a rectangular grid mesh network in Appendix \ref{sec:meshexamples}. 

While we have ignored adjusting the final output phase shifts ($\boldsymbol{\gamma}$ in Sec. \ref{sec:photonicmesh}) in Alg. \ref{alg:parallelnullification}, we emphasize that such phase shifts merely serve to define an ``output phase reference,'' which may be unimportant in some specific applications (e.g., in the optical neural network application we now discuss) but is anyway straightforward to adjust after parallel nullification of all columns.

\section{Optical neural networks} \label{sec:apps}

\begin{figure}[h!]
    \centering
    \includegraphics[width=0.48\textwidth]{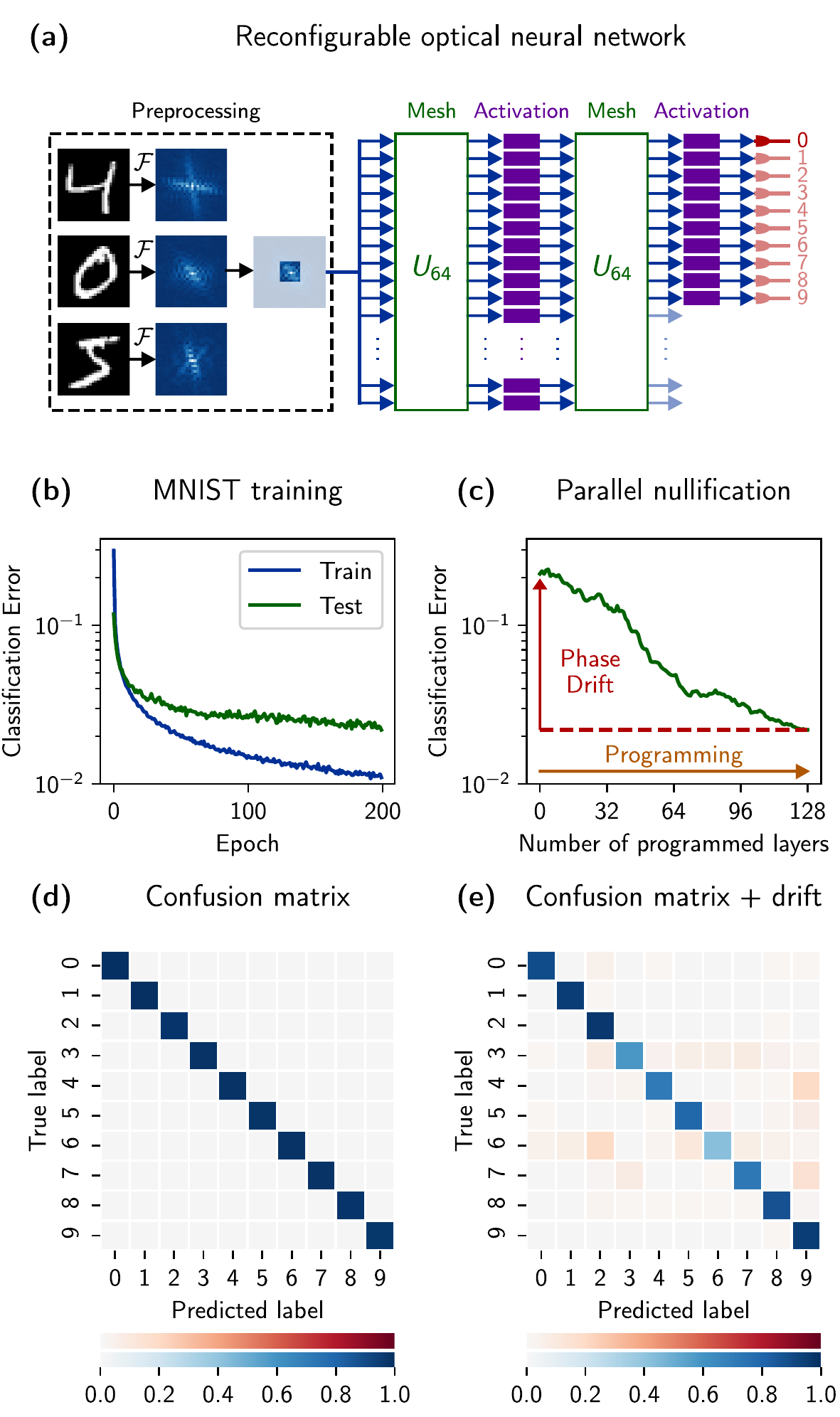}
    \caption{(a) MNIST data is preprocessed and fed as input modes into the two-layer reconfigurable neural network of Ref. \citenum{Williamson2020ReprogrammableNetworks}, and the light in the network is directed mostly towards the highlighted port corresponding to the correct label once trained. (b) Train and test accuracies for MNIST task for Adam gradient descent optimization over 200 epochs. (c) Parallel nullification corrects significant drift in phase shifter values (represented by Gaussian noise with standard deviation $\sigma_\theta = \sigma_\phi = 0.05$) and improves MNIST test accuracy. The starting classification error of 78\% corresponds to the confusion matrix with drift in (e). (d)-(e) MNIST confusion matrix for a correctly programmed ONN (d) versus with the drift (e), where colors denote predicted label percentage match to true (blue) versus incorrect (red) labels.}
    \label{fig:nn}
\end{figure}

In this section, we primarily discuss applying parallel nullification to programming reconfigurable optical neural networks (ONNs), as shown in Fig. \ref{fig:nn}. Parallel nullification can be used to diagnose and error-correct integrated optical neural networks, which are capable of performing machine learning tasks that transform optically-encoded data and can be significantly more energy-efficient than their electrical counterparts \cite{Shen2017DeepCircuits}.

For our demonstration, we choose the MNIST classification task, a popular standard in machine learning models consisting of $28 \times 28$ images of handwritten decimal digits from $0$ to $9$. Using \href{https://github.com/solgaardlab/neurophox}{\texttt{neurophox}} \cite{Pai2019MatrixDevices} and GPU-accelerated automatic differentiation in \texttt{tensorflow} \cite{Abadi2016TensorFlow:Learning}, we train a two-layer ONN (shown in Fig. \ref{fig:nn}(a)) consisting of two $N = 64$ rectangular grid meshes followed by ReLU-like optical nonlinearity or activation layers \cite{Williamson2020ReprogrammableNetworks} to classify each image to the appropriate digit label. Our training examples consist of low-frequency FFT features preprocessed from each image that are input into the device, and our ultimate goal is to direct the light into port $n + 1$ labelled by digit $n$, as demonstrated in Fig. \ref{fig:nn}(a) for digit 0. Further details on data preprocessing, model choice, and neural network robustness and performance are discussed in Appendix \ref{sec:nn} and in Ref. \citenum{Williamson2020ReprogrammableNetworks}. 

Reprogrammable electro-optic nonlinearities \cite{Williamson2020ReprogrammableNetworks} (or generally any ReLU-like optical nonlinearities that can be tuned to operate in a linear regime) allow multi-layer, ``deep'' ONNs to be programmed or calibrated using our parallel approaches. Crucially, this means that it is possible to calculate inputs to the entire ONN (rather than each layer) and sequentially program the columns of all mesh networks in the overall device to program any desired operator of choice. We now demonstrate the use of this protocol for correcting significant drifts in our specific simulated ONN.

The training of the ONN parameters is shown in Fig. \ref{fig:nn}(b), achieving $98.9\%$ accuracy on training data (60000 training examples) and $97.8\%$ accuracy on hold-out evaluation data (10000 testing examples). In our simulated environment, we use parallel nullification to correct phase drift $\delta \theta_{m, \ell}, \delta \phi_{m, \ell} \sim \mathcal{N}(0, \sigma^2)$ for $\sigma = 0.05$ in our MNIST-trained network as shown in Fig. \ref{fig:nn}(c). In particular, the confusion matrix (representing correct predictions on the diagonal and incorrect predictions off the diagonal) improves from Fig. \ref{fig:nn}(e) to Fig. \ref{fig:nn}(d) using parallel nullification, improving the test set accuracy from $78\%$ to $97.8\%$. It is important to note (and this is discussed further in Appendix \ref{sec:nn}) that only very minor decrease in performance is seen for $\sigma \leq 0.02$, which can generally be thought of as the phase shift tolerance threshold for nodes in our ONN for the MNIST task.

Parallel nullification is therefore a promising option for realizing machine learning models on reconfigurable devices \cite{Shen2017DeepCircuits, Harris2018LinearProcessors}. Such devices provide strictly more generality and flexibility over non-reconfigurable ONNs which can only implement one model (specified pre-fabrication \cite{Fang2019DesignImprecisions}) and furthermore cannot be dynamically error-corrected post-fabrication.

\section{Error considerations} \label{sec:error}

With our optical neural network example, we have shown how parallel nullification can correct phase drift and thus improve performance considerably. We further discuss the fault tolerance of parallel nullification to sources of systematic error that arise during fabrication of feedforward mesh networks.

\subsection{Phase errors}

The parallel nullification step in Eq. \ref{eqn:nullification} is agnostic to any static phase shifts that may accumulate at each column due to path length variations, which in other cases (e.g., non-reconfigurable systems) result in phase errors. Specifically, parallel nullification implicitly sets the reference by which phases in the device are measured \cite{Miller2017SettingMethod}, so the nullification set calculation of Eq. \ref{eqn:nullvector} gives the correct inputs for parallel nullification regardless of how the node is controlled. A correctly programmed mesh can be achieved using a TDC- or MZI-based tunable beamsplitter as depicted in Fig. \ref{fig:parallelnullification}(d) or any of the variations discussed in Appendix \ref{sec:tb}. 

\subsection{Split ratio error}

Parallel nullification can correct split ratio errors similarly to how phase errors are corrected, but in some cases, the split ratio range can be limited at each node of the photonic mesh (e.g., due to imperfect 50/50 beamsplitters in typical MZIs \cite{Miller2015PerfectComponents}). This limited range problem can be avoided entirely using TDCs or double-MZIs \cite{Miller2015PerfectComponents} rather than single MZIs at each node. As discussed in Appendix \ref{sec:tb}, we can equivalently consider $\theta$ as the ``tunable coupling constant'' for the TDC. A TDC can achieve perfect operation, or full split ratio range, if the modes are phase matched over the entire tunable range of $\theta$. We could ensure this range is achievable by making the device suitably long such that the full split ratio range is contained between the minimum and maximum extent of the tunable coupling constant (i.e., such that $\theta \in [0, \pi]$).

\subsection{Thermal crosstalk}

Thermal crosstalk between device elements can occur whenever there is significant heat generated in the phase-shifting process, such as in thermal phase shifters. We assume that thermal crosstalk is very small between columns and only occurs within each column so that calibrations of past columns are not affected by those of future columns. As this thermal crosstalk increases, the parallel nullification of each column takes longer because the optimizations within each column are no longer independent of each other (e.g., the settings of node $m$ would be affected by the settings of nodes $m - 1$ and $m + 1$). If running parallel nullification, however, we might still be able to efficiently find an optimal setting for the column as long as this crosstalk is reasonably small. Furthermore, phase shifter technologies that have little to no crosstalk (such as MEMS phase shifters \cite{Edinger2019Low-lossPhotonics}) would be faster to program because nullifications within the column would be truly independent. In Appendix \ref{sec:tb}, we propose node configurations with at most $\pi$ phase shifts (rather than $2\pi$), requiring smaller temperature variations per waveguide length and limiting thermal crosstalk compared to conventional designs.

\subsection{Loss}

Parallel nullification is capable of programming loss-balanced architectures. A loss-balanced architecture is achieved if all the waveguide path lengths and bends are equal (assuming uniform waveguide scattering loss) and the phase shifters are lossless (i.e., changing a phase shift does not increase or decrease loss incurred by that phase shift). If all modes encounter a loss $\mu_\ell$ at each column $\ell$, then it is straightforward to show the column can be programmed to implement $\mu_\ell U^{(\ell)}_N$ using parallel nullification. From Eq. \ref{eqn:ffmesh}, the overall network implements $\mu U_N$, where $\mu$ is ideally a ``global loss'' equal to the product of all column-wise losses, i.e. $\mu = \prod_\ell \mu_\ell$ \cite{Harris2018LinearProcessors}. Loss-balanced grid architectures (such as rectangular grid meshes \cite{Clements2016AnInterferometers}) and other symmetric architectures such as the butterfly (FFT) architecture \cite{Flamini2017BenchmarkingProcessing} can be fabricated to fit this criterion. Other architectures (e.g., triangular meshes \cite{Miller2013Self-configuringInvited}) can include ``dummy'' elements to achieve the same balance. In the case of optical neural networks, some additional calibration of the nonlinear elements may also be required in the presence of loss, which may benefit from reprogrammability of such elements \cite{Williamson2020ReprogrammableNetworks}.

If the feedforward mesh (or specifically a column of the mesh) suffers from ``loss imbalance,'' then different amounts of light are lost from each output as light propagates. In this case, we might need to readjust the nullification set (e.g., by adjusting the computer model of the mesh to account for lossy mesh columns) to more accurately program in the desired operator, which is a direction that should be further explored.

\section{Discussion and Conclusion}

We derive a graph-topological property for any reconfigurable feedforward photonic network of tunable beamsplitter nodes that allows efficient programming (``parallel nullification'') of node columns that are not affected by each other and thus can be tuned simultaneously. With a model of the device stored in a computer, we find a set of vector inputs to the device (the ``nullification set'') and for each column, nullify the bottom power of all tunable beamsplitters in parallel using the corresponding nullification vector. The nullification set can be internally generated given a single-mode input by appending the optical setup machine of Ref. \citenum{Miller2017SettingMethod} to the mesh.

Parallel nullification can quickly diagnose and error-correct phase drifts of a reconfigurable photonic device, demonstrated in the context of reconfigurable optical neural networks in Section \ref{sec:apps}. As nullification set inputs are sent in order from $\ell = 1$ to $\ell = L$, error appears as non-nullified power at the bottom output ports of the problematic column. This error may be corrected by nullifying these ports so that further debugging of the photonic circuit can be performed.

Our programming algorithm differs from calibration schemes \cite{Harris2017QuantumProcessor, Shen2017DeepCircuits, Carolan2015UniversalOptics, Mower2015High-fidelityCircuits} that fully characterize all tunable elements (e.g. relating phase shift to voltage via a cubic model). Such approaches can achieve high fidelities, but components that experience environmental drift need to be recalibrated. In contrast, parallel nullification is not tied to any specific calibration model and should readily adapt to calibration drift and environmental perturbations. 

However, as suggested in the Introduction, it is possible to apply our graph-topological arguments to ``parallel calibration.'' We explicitly provide this parallel calibration protocol in Appendix \ref{sec:parallelcalibration} that is similar in principle to current calibration protocols \cite{Harris2017QuantumProcessor, Shen2017DeepCircuits, Carolan2015UniversalOptics, Mower2015High-fidelityCircuits}, but with notable differences (e.g., increased efficiency via parallelization) and simplifications (e.g., removing the need to explicitly calibrate ``meta-MZI'' cells \cite{Harris2017QuantumProcessor}). At a high level, each node column is calibrated in parallel, assuming no crosstalk among elements in the column, by simultaneously sweeping phase shifter values and measuring corresponding powers in embedded detectors. Such calibration can elucidate phase shift-voltage relationships, which are required for initialization of the network or \textit{in situ} backpropagation with respect to the actual node voltages \cite{Hughes2018TrainingMeasurement}, which can be useful in the context of training optical neural networks.

Our approach is similar to the RELLIM approach of Ref. \citenum{Miller2017SettingMethod}, where each input tunes a \textit{single} node, since it relies on the reciprocity of linear optical networks. However, each input in parallel nullification tunes an entire \textit{column} of nodes simultaneously based on the feedforward mesh definition proposed in Eq. \ref{eqn:ffmesh}. Our approach can be fault-tolerant for feedforward meshes with phase shift crosstalk (i.e., thermal crosstalk \cite{Shen2017DeepCircuits}) and beamsplitter fabrication errors. Additionally, parallel nullification is currently the most efficient protocol for programming or calibrating a feedforward photonic mesh. In particular, where $L$ is the number of device columns and $N$ is the number of input modes, our parallel nullification protocol requires just $L$ input vectors and programming steps, resulting in up to $(N / 2)$-times speedup over existing component-wise calibration approaches.

\section*{Funding Information}
Air Force Office of Scientific Research (AFOSR), specifically for the Center for Energy-Efficient 3D Neuromorphic Nanocomputing (CEE3N$^2$) and a MURI program, Grant Nos. FA9550-18-1-0186 and FA9550-17-1-0002 respectively.

\section*{Acknowledgments}

We would like to thank Nathnael Abebe, Ben Bartlett, and Rebecca L Hwang for useful discussions.


\bibliography{parallelnull}

\appendix

\section{Neurophox: open source software} \label{sec:neurophox}
In our simulation framework \href{https://github.com/solgaardlab/neurophox}{\texttt{neurophox}}, we provide our general definition of feedforward mesh architectures from Eq. \ref{eqn:ffmesh} and the reconfigurable neural network model of Fig. \ref{fig:nn}. This is the first time to our knowledge that feedforward meshes have been defined in this way, and it allows for a greatly simplified framework for defining and simulating mesh architectures. In \href{https://github.com/solgaardlab/neurophox}{\texttt{neurophox}}, we provide Python code to calculate the nullification set and simulate parallel nullification on a physical chip using this nullification set. We also provide the code to train fully optical feedforward neural networks on the MNIST dataset with automatic differentiation in \texttt{tensorflow}.

\section{Two-step nullification}\label{sec:nullificationproof}

We give an explicit proof of the two-step nullification protocol in Ref. \citenum{Miller2013Self-configuringInvited} in our context. In our setup for this proof, we have an input vector $\boldsymbol{u} = (u_1, u_2)$ to a node $T_2$, yielding output vector $\boldsymbol{x} = (x_1, x_2)$, i.e. $\boldsymbol{x} = T_2(\alpha, \beta) \boldsymbol{u}$ (defined in Eq. \ref{eqn:tunablebeamsplitter} of the main text). We would like to find the $\alpha = \theta, \beta = \phi$ such that $x_2 = 0$.

In particular, we want to show that minimizing bottom port power ($|x_2|^2$) with respect to $\beta$ gives $\beta = \phi$ regardless of the value of $\alpha$. Then, it is clear that if we sweep $\alpha$ to nullify power, we must have $\alpha = \theta$.
\begin{equation}
\begin{aligned}
x_2 &= e^{i \beta}\cos \frac{\alpha}{2} u_{1} - \sin \frac{\alpha}{2} u_{2}\\
|x_2|^2 &= \cos^2 \frac{\alpha}{2} |u_1|^2 + \sin^2 \frac{\alpha}{2} |u_2|^2 \\& \hspace{-0.1cm}- 2 \cos \frac{\alpha}{2} \sin \frac{\alpha}{2}|u_1||u_2| \mathrm{Re}(e^{i\mathrm{arg}(u_1)}e^{-i\mathrm{arg}(u_2)}e^{i\beta})\\
\beta^{\mathrm{opt}} &:= \displaystyle{\min_{\beta \in [0, 2\pi)} |x_2|^2} = -\arg\left(\frac{u_1}{u_2}\right) = \phi,
\end{aligned}
\end{equation}
where we allow any $\alpha \in [0, \pi]$, i.e. $\cos \frac{\alpha}{2} \sin \frac{\alpha}{2} \geq 0$. Now that we have optimized $\beta$, we optimize $\alpha$ to completely nullify $|x_2|^2$.
\begin{equation}
\begin{aligned}
|x_2|^2 &= \left(\cos \frac{\alpha}{2} |u_1| - \sin \frac{\alpha}{2} |u_2|\right)^2 \stackrel{?}{=} 0\\
\alpha^{\mathrm{opt}} &= 2\arctan\left|\frac{u_1}{u_2}\right| = \theta,
\end{aligned}
\end{equation}
These results match the desired expressions of Eq. \ref{eqn:nullification}.

\section{Feedforward mesh examples} \label{sec:meshexamples}

\begin{figure}[h]
    \centering
    \includegraphics[width=0.5\textwidth]{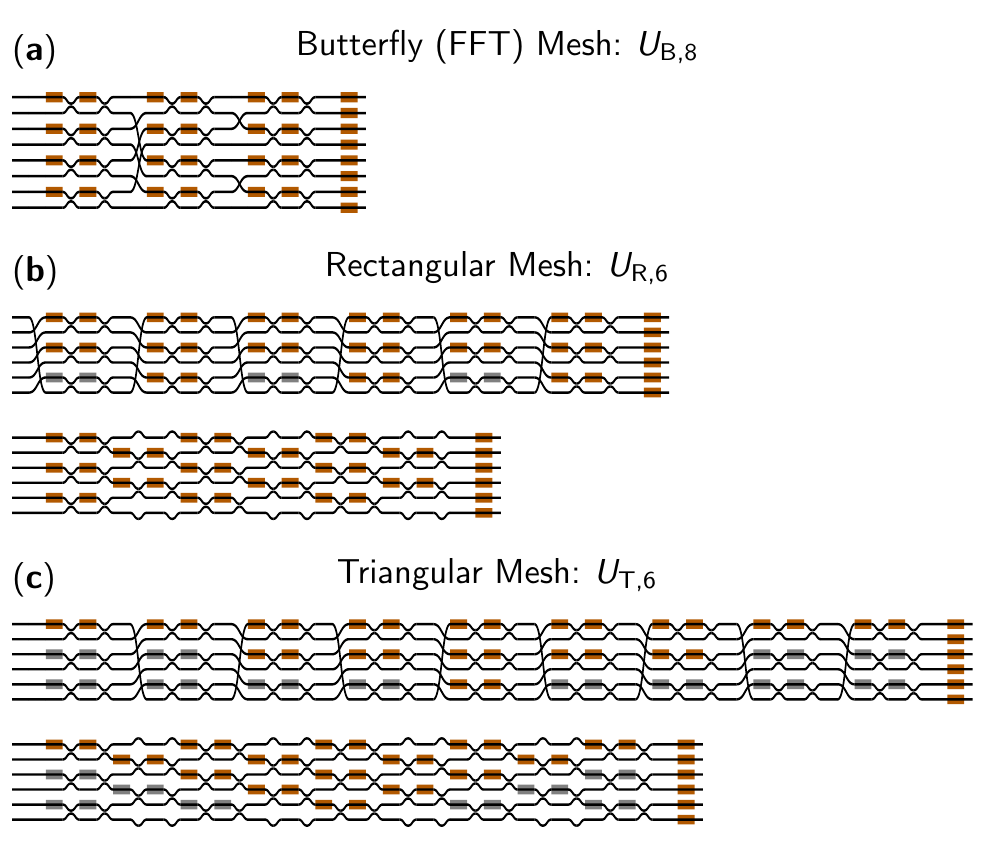}
    \caption{For illustrative purposes, we show example mesh diagrams for (a) butterfly, (b) rectangular, and (c) triangular meshes. For (b)-(c), we show (top) how the device is modelled (using the expression of Eq. \ref{eqn:ffmesh} in the main text) and (bottom) how it is physically implemented (the ``grid'' design). In all mesh diagrams, orange phase shifters represent tunable phase shifters $\boldsymbol{\theta}, \boldsymbol{\phi}, \boldsymbol{\gamma}$. Gray phase shifters represent bar state MZIs ($\theta = \phi = \pi$).}
    \label{fig:meshexamples}
\end{figure}

\begin{figure}[h]
    \centering
    \includegraphics[width=0.48\textwidth]{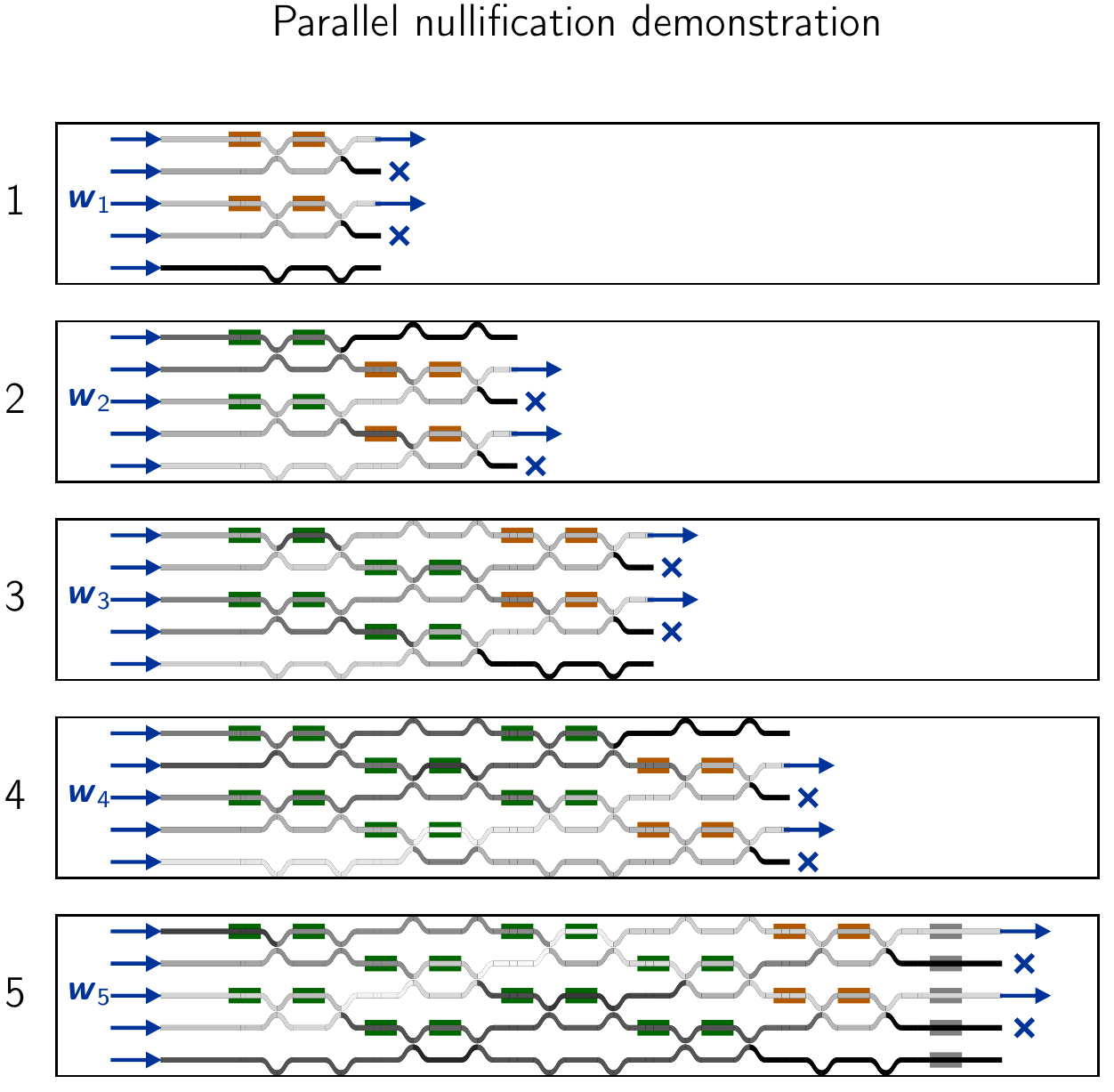}
    \caption{Parallel nullification of Fig. \ref{fig:parallelnullification} simulated on an $N = 5$ feedforward rectangular grid mesh in \href{https://github.com/solgaardlab/neurophox}{\texttt{neurophox}}. The relative field magnitudes are represented by grayscale values, the $\theta_{m, \ell}, \phi_{m, \ell}$ phase shifts are green and the $\gamma_n$ are gray. At each step denoted on the left, we tune each column $\ell$ (depicted in orange) in parallel given input $\boldsymbol{w}_\ell$. This is accomplished progressively (indicated by the black-colored waveguides and blue crosses) from left to right until the full matrix is tuned.}
    \label{fig:rm}
\end{figure}

We can apply the general definition of a feedforward mesh to commonly studied architectures shown in Fig. \ref{fig:meshexamples}. Grid architectures include the rectangular or Clements mesh \cite{Clements2016AnInterferometers} and the triangular or Reck mesh \cite{Reck1994ExperimentalOperator}, which are both universal photonic mesh architectures. For a rectangular mesh, each of the $L = N$ columns has $M_\ell = M - 1 + \ell (\mathrm{mod}\ 2)$ and $P_\ell$ defined as an upward circular shift for odd $\ell$ and a downward circular shift for even $\ell$. For a triangular mesh, each of the $L = 2N - 3$ columns has $M_\ell = \lceil\frac{\min(\ell, 2N - 2 - \ell)}{2}\rceil$ with the same $P_\ell$ as the rectangular mesh. The fact that the triangular mesh has the same $P_\ell$ as the rectangular mesh allows it to be ``embeddable'' within a rectangular mesh.

As shown in Fig. \ref{fig:rm}, a physical rectangular grid mesh (note the graph-topological transformation from Fig. \ref{fig:meshexamples}) can be progressively tuned to implement any unitary matrix by performing parallel nullification protocol.

The butterfly architecture is an example of an alternative feedforward architecture that can be tuned using parallel nullification. Such architectures are designed to be compact, robust, and fault-tolerant alternatives to universal meshes \cite{Flamini2017BenchmarkingProcessing, Fang2019DesignImprecisions}. This means that implementing reconfigurable architectures of this form is potentially more scalable (to the feature size or number of inputs and outputs) and can be useful for machine learning approaches even though it cannot actually implement any arbitrary unitary operator. The operations that the butterfly mesh \textit{can} implement, however, is the discrete Fourier transform (DFT) operator and (when concatenated with its mirror image to form the Benes network) any permutation operator.

Even meshes that are not explicitly feedforward-only \cite{Perez2017HexagonalInterferometers, Perez2017MultipurposeCore}, but are meant for more general-purpose approaches, can in theory implement the rectangular or triangular grid architectures. With some additional characterization, it may be possible to program or calibrate such general architectures in the lab setting using our method.
\begin{figure*}[t!]
    \centering
    \includegraphics[width=0.4\linewidth]{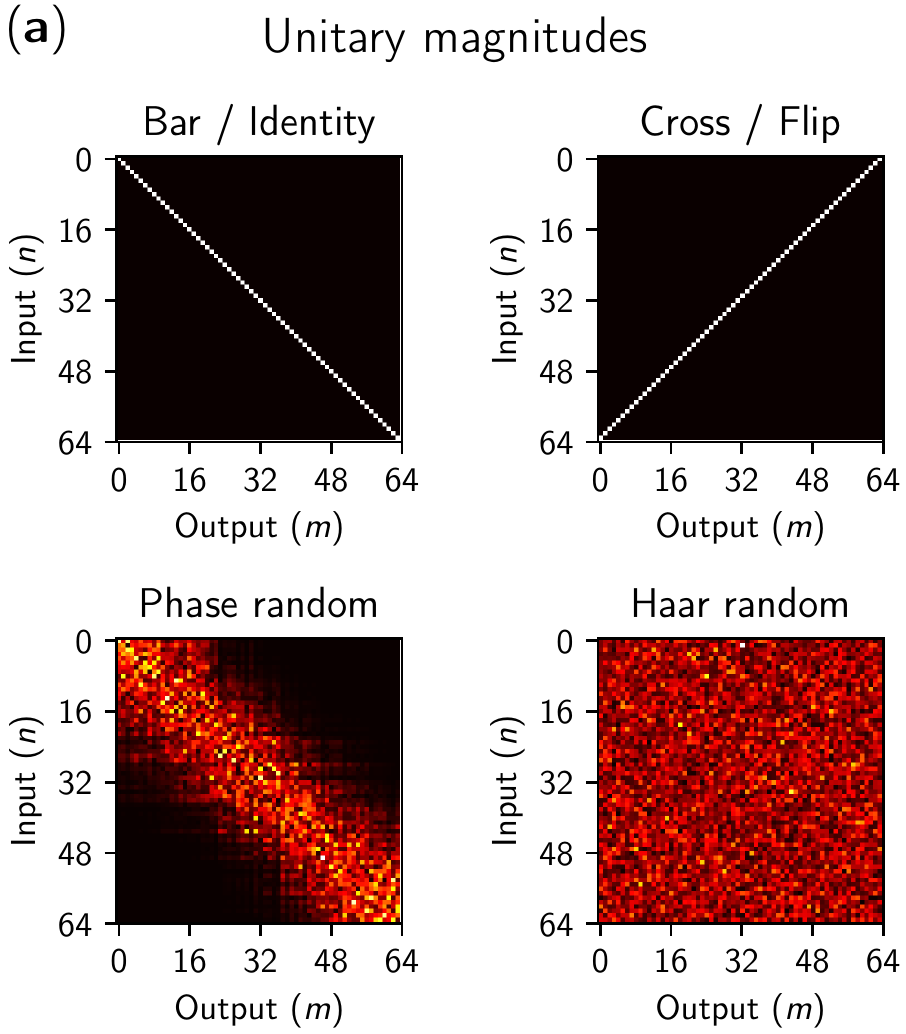}
    \hspace{1cm}
    \includegraphics[width=0.4\linewidth]{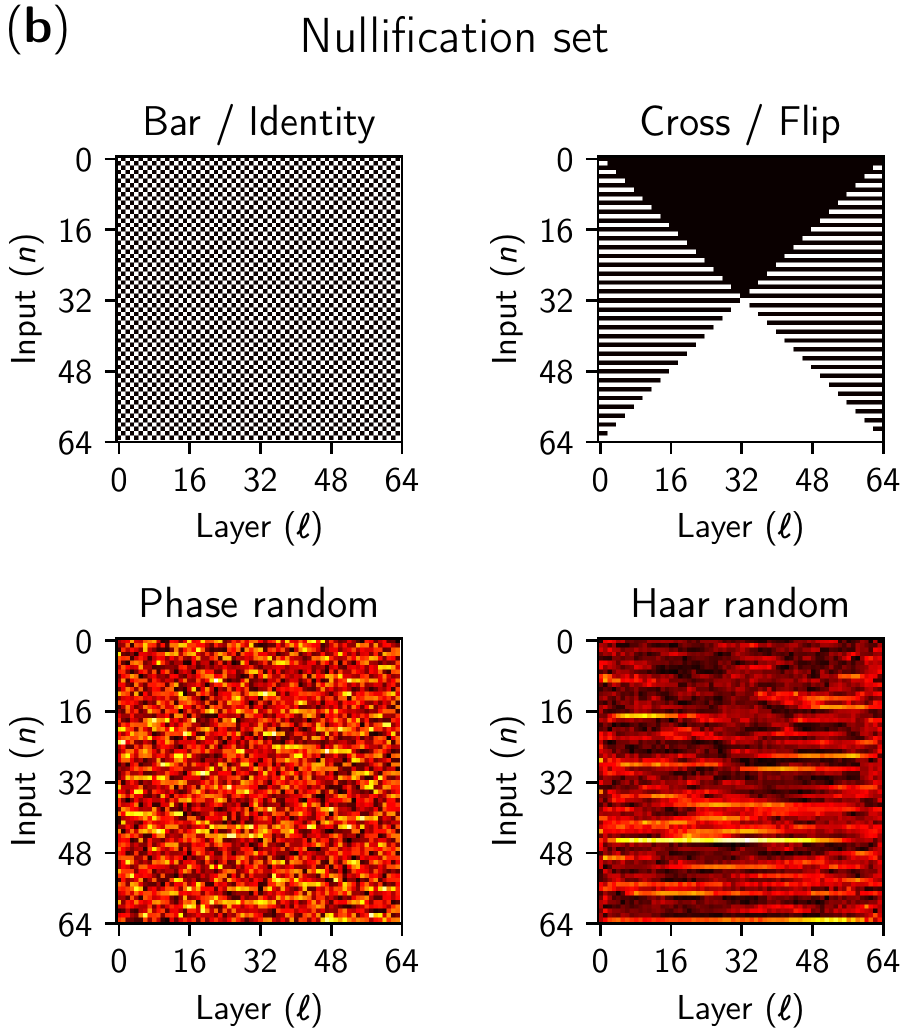}
    \caption{(a) Magnitudes of the unitary matrix elements for a rectangular mesh for size $N = 64$ given bar state, cross state, uniformly random phase (``phase random''), and random unitary (``Haar random'') phase settings. (b) Power magnitudes of the nullification set $\boldsymbol{w}_\ell$ for a rectangular mesh for size $N = 64$ given bar state, cross state, uniformly random phase, and Haar random phase settings.}
    \label{fig:nullificationset}
\end{figure*}

\section{Nullification set patterns} \label{sec:nullset}

While the nullification set is mostly useful in the progressive calibration of columned optical meshes, there is interesting structure in the nullification vectors for a random unitary matrix implemented on a rectangular mesh device.

For a rectangular mesh, the implementation of these Haar random unitary matrices (i.e. matrices with roughly uniform-random magnitude elements) for large $N$ involves low reflectivity in the center of the mesh and random reflectivity on the boundary \cite{Russell2017DirectMatrices, Pai2019MatrixDevices}. For our choice of normalization basis calculation in Eq. \ref{eqn:nullvector}, the nullification vectors of a Haar random matrix lie somewhere between those for a mesh of only bar state nodes (as indicated by the ``bar/identity'' label) and those of a mesh of only cross state nodes (as indicated by the ``cross/flip'' label) in Fig. \ref{fig:nullificationset}. As expected, the nullification set for the cross/flip mesh has an antisymmetric configuration versus the more symmetric configuration of the nullification set for the bar/identity mesh in Fig. \ref{fig:nullificationset}. We note that the vectors $\boldsymbol{w}_\ell$ are not generally mutually orthogonal, so the nullification set also generally do not form a unitary matrix if arranged side by side.

There is also an information theoretic connotation of the nullification set. The nullification set for a mesh with phase settings that are uniformly set from $[0, 2\pi)$ (which leads to the ``banded unitary'' in Fig. \ref{fig:nullificationset}(a)) looks more random (i.e. less structured) than the nullification set for the Haar-random unitary (the truly random unitary in Fig. \ref{fig:nullificationset}(a)). This is due to the nonlinear relationship between the transmission amplitude of the individual nodes and the final output magnitudes of the overall device \cite{Russell2017DirectMatrices, Pai2019MatrixDevices}.

\begin{figure}[h]
    \centering
    \includegraphics[width=\linewidth]{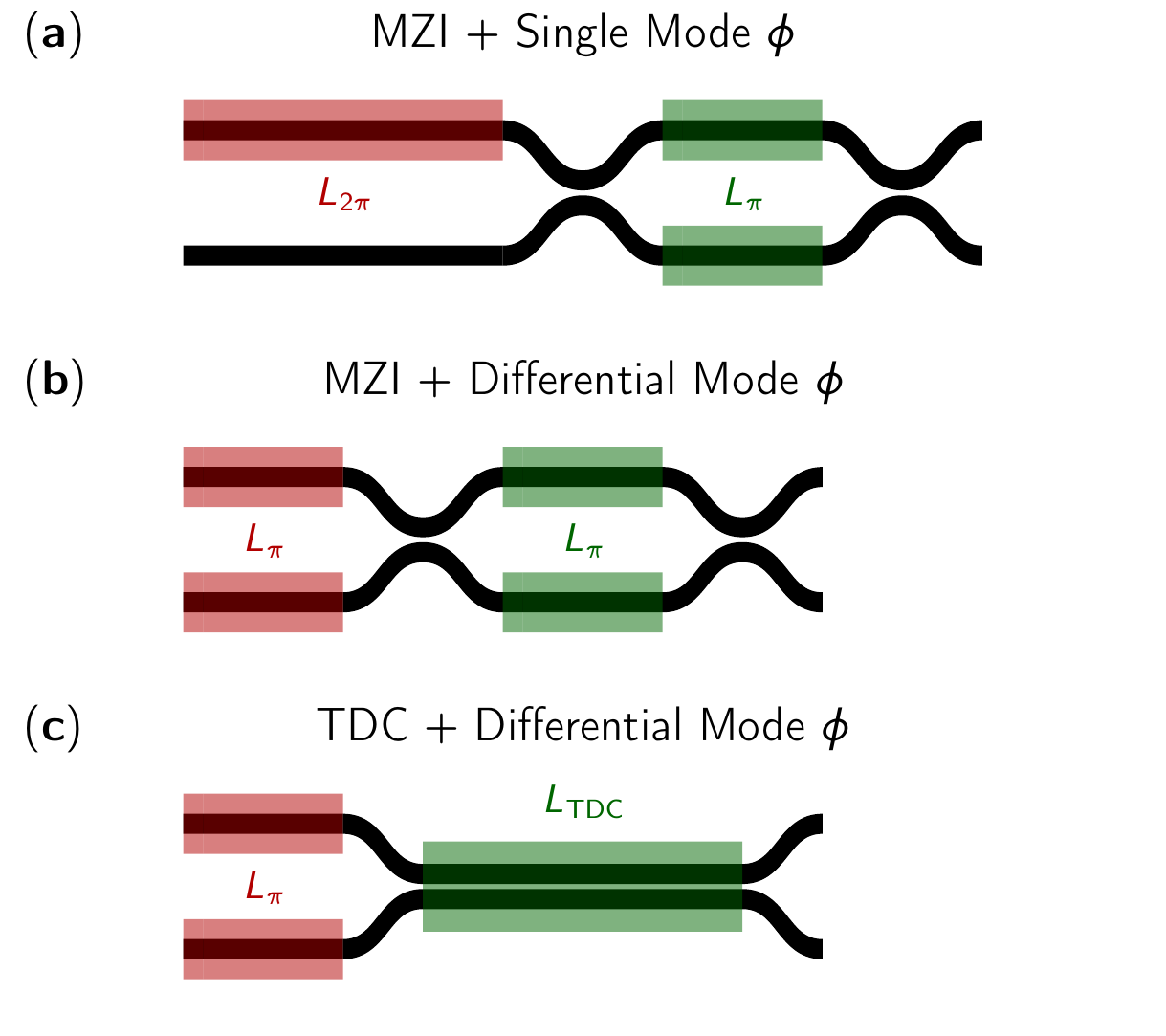}
    \caption{Options for tunable beamsplitter $T_2(\theta, \phi)$, where $\theta$ is controlled by the green block and $\phi$ is controlled by the red block. (a) the usual non-compact configuration; (b) the compact differential-mode MZI; (c) the compact TDC.}
    \label{fig:mzi}
\end{figure}

\section{Tunable beamsplitter variations} \label{sec:tb}

There are many possible ways of positioning phase shifters in an MZI that are all ultimately equivalent in allowing universal unitary meshes and parallel nullification. The simplest general statement is that, for a $2 \times 2$ MZI node, we need two phase shifters, one of which must be on one waveguide arm inside the MZI to control the split ratio. The second phase shifter can be on any input or output waveguide \cite{Reck1994ExperimentalOperator} or on the other waveguide arm inside the MZI \cite{Miller2013Self-configuringInvited}. If the second phase shifter is on an input arm of the MZI, then we need additional phase shifters for the mesh to implement an arbitrary unitary operator. Though such a design is well-suited for self-configuration and setting up input vectors into the device, the protocol for parallel nullification may be less straightforward since such this symmetric configuration does not have an equivalent form as in Eq. \ref{eqn:tunablebeamsplitter}. For simplicity in programming the device, we primarily concern ourselves with configurations where the second phase shifter is on the \textit{input} waveguide obeying the form of Eq. \ref{eqn:tunablebeamsplitter}.

A commonly proposed alternative to the MZI for the split ratio modulation ($S_\mathrm{A}$ in Eq. \ref{eqn:tunablebeamsplitter} of the main text) is the tunable directional coupler (TDC), which can achieve any split ratio by simply tuning the coupling region of a directional coupler. The transmissivity varies as $t = \cos^2(\kappa L_{\mathrm{TDC}})$ with $\theta = \kappa L_{\mathrm{TDC}}$ and $\kappa$ the tunable coupling constant from coupled mode theory. Phase matched modes should always allow for the full range of transmissivities as long as the range of $\kappa L_{\mathrm{TDC}}$ is large enough. One proposal that follows the TDC scheme is that of Ref. \citenum{Miller2019PhaseMovement}.

An alternative control scheme for phase shift operator $S_\mathrm{P}$ in Eq. \ref{eqn:tunablebeamsplitter} of the main text is a ``differential mode'' phase shifter scheme. To make meshes more compact, it is functionally equivalent to have phase shifters in both the top and bottom waveguides that can reach a maximum of $\pi$ phase shift. Tuning $\phi$ in the range of $[0, \pi)$ would then consist of tuning the top phase shifter from steady state, whereas tuning $\phi$ in the range of $[\pi, 2\pi)$ would consist of tuning the bottom phase shifter from steady state. Of course, the tradeoff of this more compact scheme is increased complexity in the number of electrical contacts and the logic of the nullification protocol. Independent of any of these control schemes, the transmission matrix model of Eq. \ref{eqn:tunablebeamsplitter} generates the correct set of nullification vectors, so parallel nullification works for any of these schemes.

\section{Neural network training} \label{sec:nn}

\begin{figure}
    \centering
    \includegraphics[width=0.48\textwidth]{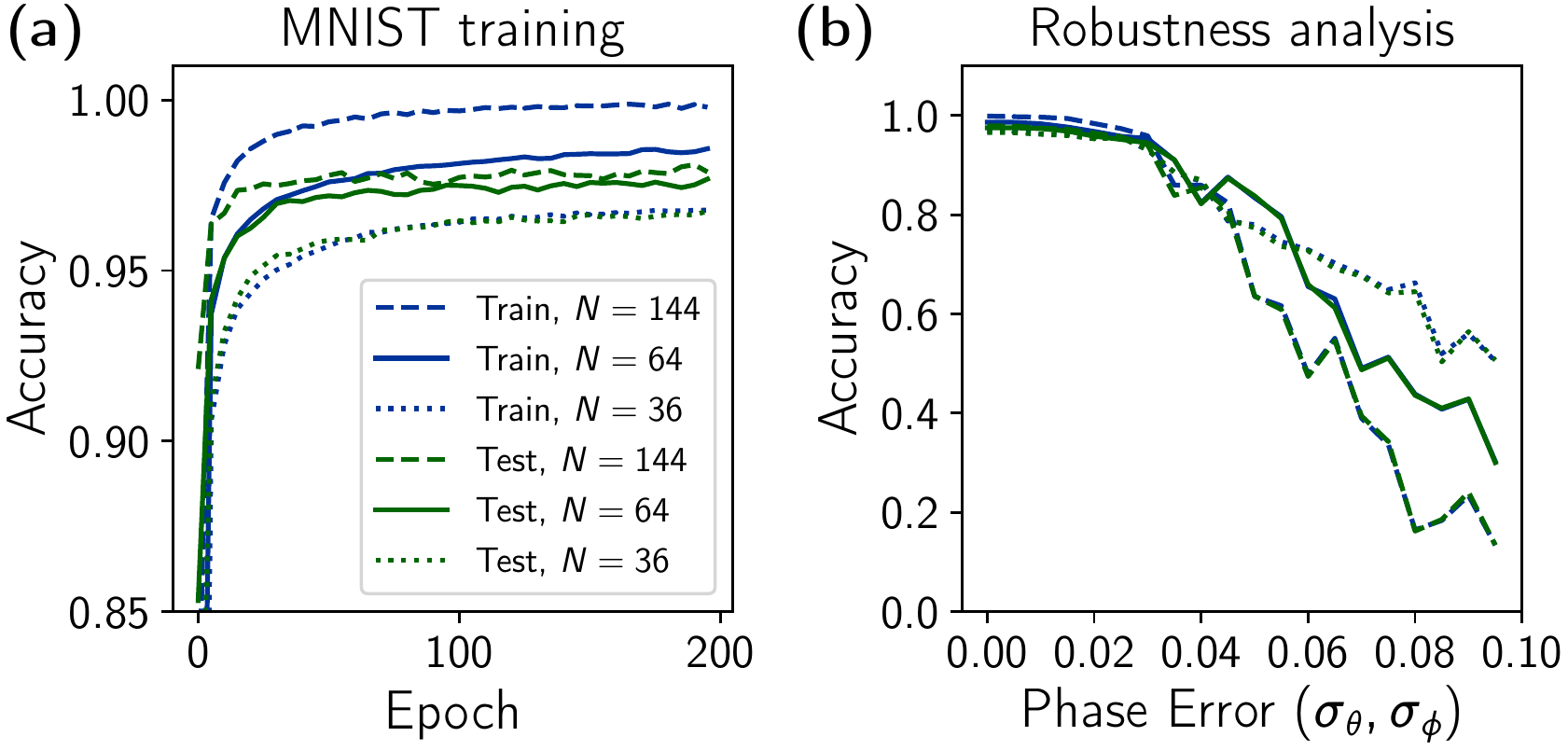}
    \caption{(a) MNIST training results for $N = 36, 64, 144$ for the two-column neural network depicted in Fig. \ref{fig:nn}(a). Significant overfitting is observed when $N = 144$ due to lack of regularization. (b) Accuracy robustness analysis for the ONN for $N = 36, 64, 144$ for phase errors ranging from $\sigma = 0$ to $\sigma = 0.1$.}
    \label{fig:nnanalysis}
\end{figure}

As in Ref. \citenum{Williamson2020ReprogrammableNetworks}, we preprocess each image by applying a Fourier transform and discrete low-pass filter, since such a task is potentially feasible in the optical domain via Fourier optics. In our low-pass feature selection, we pick the center $8 \times 8$ block of pixels since most of the useful data is within this low-frequency band, giving us a total of $64$ features. Unlike in Ref. \citenum{Williamson2020ReprogrammableNetworks}, we retain some of the redundant Fourier features in this window since it slightly boosts neural network performance, and we use a mean square error loss instead of a categorical cross entropy loss due to slight performance improvement in the former.

Our prediction consists of dropping the final $54$ outputs of the second neural network layer (shown by the light blue arrows in Figure \ref{fig:nn} of the main text), and squaring the amplitudes of the remaining $10$ outputs (equivalent to taking a power measurement, denoted by the red photodetector symbols of Figure \ref{fig:nn}). Given the $d$th data sample $(\boldsymbol{x}_d, \boldsymbol{y}_d)$ (pair of input feature vector and one-hot label vector), our cost function $\mathcal{L}$ is the mean square error
\begin{equation} \label{sec:mse}
    \mathcal{L} = \sum_{d = 1}^{D}\left\lVert \frac{f(\boldsymbol{x}_d)}{\|f(\boldsymbol{x}_d)\|} - \boldsymbol{y}_d \right\rVert^2,
\end{equation}
where $f(\boldsymbol{x}_d)$ represents the raw output powers of the neural network given input $\boldsymbol{x}_d$. In practice, our classification will always correspond to the port in which the highest output power is measured, ideally guiding input mode vector $\boldsymbol{x}_d$ to the output port corresponding to $\boldsymbol{y}_d$. Our model achieves a final train accuracy of $98.9\%$ and a final test accuracy of $97.8\%$. Slightly worse training performance was found using a categorical cross entropy loss.

Finally, we perform an robustness analysis of imperfections in neural network performance (as in Ref. \citenum{Fang2019DesignImprecisions}). We train the same two-column neural network of Figure \ref{fig:nn} for different sizes of $N = \{36, 64, 144\}$, achieving test accuracies of $96.6\%, 97.8\%, 98.1\%$ respectively, though performance varies slightly from run to run. The training curves and robustness analysis for different phase errors (Gaussian phase errors of the form $\mathcal{N}(0, \sigma^2)$) are shown in Figure \ref{fig:nnanalysis}, where we find that phase errors above $\sigma = 0.02$ begin to significantly affect performance. Of course, once parallel nullification is applied, any such errors can be corrected regardless of the exact calibration model for the individual phase shifters. Note that in the case $N = 144$, significant overfitting is present, though adding dropout (i.e., zeroing-out power in some of the ports) after the first ONN layer might be one method to ``regularize'' the physical ONN and therefore reduce this overfitting.

We note that the study in Ref. \citenum{Fang2019DesignImprecisions} studies the MNIST task on larger simulated ONNs and accomplishes a similar test accuracy ($97.8\%$). Aside from training significantly fewer parameters than in Ref. \citenum{Fang2019DesignImprecisions}, we use unitary rather than general linear layers, we use mean square error rather than categorical cross entropy loss, and our feature selection is different (low-frequency Fourier features rather than all raw features).

\section{Parallel calibration} \label{sec:parallelcalibration}

\begin{figure}[h]
    \centering
    \includegraphics[width=\linewidth]{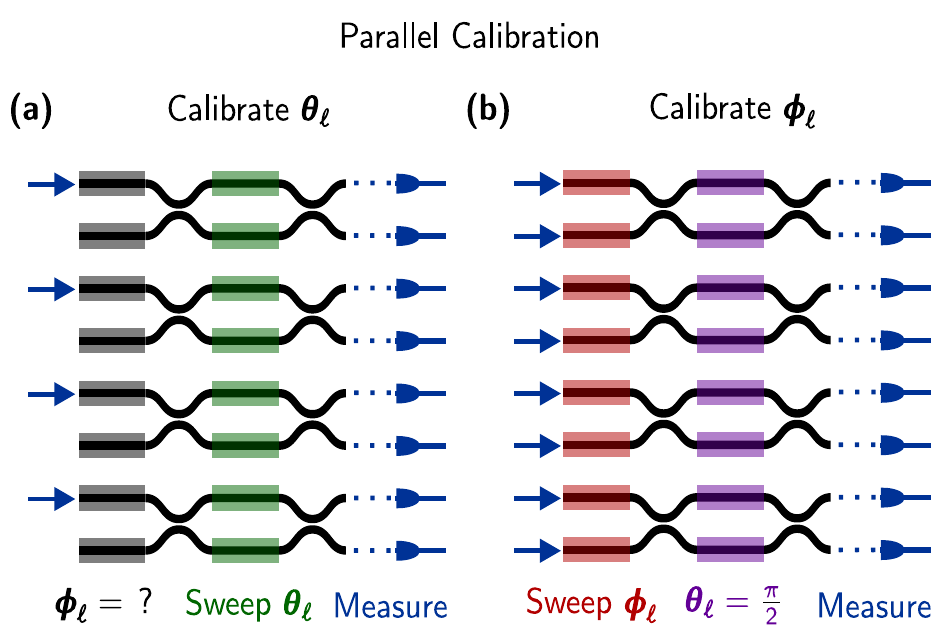}
    \caption{Parallel calibration of any node column proceeds in two steps (where all blue arrows in the figure indicate a mode with the same amplitude). In (a), we send in the appropriate input (such that $\boldsymbol{u}_\ell = (1, 0, 1, 0, \ldots)$) to calibrate $\boldsymbol{\theta}_\ell$ regardless of how $\boldsymbol{\phi}_\ell$ is calibrated. In (b), we send in the appropriate input (such that $\boldsymbol{u}_\ell = (1, 1, 1, 1, \ldots)$) to calibrate $\boldsymbol{\phi}_\ell$ given the setting $\boldsymbol{\theta}_\ell = \pi / 2$ (which we know after calibrating $\boldsymbol{\theta}_\ell$).}
    \label{fig:parallelcalibration}
\end{figure}

When calibrating a large number of nodes in a feedforward mesh \cite{Harris2017QuantumProcessor, Carolan2015UniversalOptics, Taballione20188x8Waveguides}, it is generally convenient to have a fast method to generate calibration curves for the voltage drive of each tunable element in the device. Like parallel nullification, our ``parallel calibration'' protocol proceeds from left-to-right and can generate these transmission curves in parallel for all nodes within a given column.

In some grid meshes, all nodes can be systematically tuned to cross state (e.g., in the programmable nanophotonic processor (PNP) \cite{Harris2017QuantumProcessor} or self-configuring triangular grid network \cite{Miller2015PerfectComponents}). In such schemes, power maximization at the appropriate output detectors can be used to remove the need for embedded detectors (i.e., use output detectors for the entire calibration procedure). 

Other feedforward schemes, however, require embedding photodetectors for parallel nullification, which may as well be used for parallel calibration. To calibrate $\boldsymbol{\theta}_\ell$, we find the necessary input vector so that the bottom input port of all nodes in column $\ell$ are nullified (i.e., input power of $(1, 0)$ to all nodes in the column) while setting all previously calibrated layers to bar state. We then measure resulting output transmissivities as we sweep all $\boldsymbol{\theta}_\ell$ simultaneously from the minimum to maximum allowable voltage drive settings, as shown in Fig. \ref{fig:parallelcalibration}(a). To calibrate $\boldsymbol{\phi}_\ell$, we input the uniform power in all mesh inputs such that all nodes in column $\ell$ have equal power in both input ports (i.e., input power of $(1, 1)$ to all nodes in the column). Assuming we have already calibrated $\boldsymbol{\theta}_\ell$, we set all tunable $\boldsymbol{\theta}_\ell = \pi / 2$. Now, we can calibrate $\boldsymbol{\phi}_\ell$ just as we calibrated $\boldsymbol{\theta}_\ell$ as shown in Fig. \ref{fig:parallelcalibration}(b), similar to a parallelized version of the meta-MZI scheme \cite{Harris2017QuantumProcessor, Mower2015High-fidelityCircuits}. This calibration approach works for any of the node configurations in Fig. \ref{fig:mzi}. As outlined in Ref. \citenum{Harris2017QuantumProcessor}, we can calibrate the phase given the transmissivity $T$ using the relation $T = \sin^2(\theta(v) + \theta_0)$, where $\theta_0$ is the reference phase and $\theta(v)$ is the calibration curve given a sweep over all possible voltage settings for $v$. Once the sweep is finished, we define our calibration by storing the full phase shift-voltage lookup table or a cubic model fit to that curve (3 parameters per tunable element or 6 parameters per node) \cite{Carolan2015UniversalOptics}.

Calibrating $\boldsymbol{\theta}_\ell$ then $\boldsymbol{\phi}_\ell$ for each layer of the mesh from left-to-right results in a fully calibrated device, irrespective of the feedforward architecture. Once calibrated, any reachable unitary operator on the device can be implemented by simply flashing desired values based on the measured transmission curves. Our parallel calibration procedure would need to be repeated once the nodes in the device experience fatigue or environmental perturbation, leading to calibration drift. As mentioned in the main text, this calibration drift might be diagnosed efficiently by sending in nullification set input vectors to identify errors within each column of the feedforward network denoted by unnullified nodes for the corresponding column. Our calibration can also be used to initialize parallel nullification or calculate the final updates for a procedure like \textit{in situ} backpropagation \cite{Hughes2018TrainingMeasurement}. Our parallel calibration, like parallel nullification, generally give us up to $O(N)$ speedup and is also generally applicable to any feedforward mesh. It is worth noting also that our parallel calibration protocol might be transferrable to existing technologies such as the PNP grid architecture, with some minor modifications of the protocol discussed in the supplementary of Ref. \citenum{Harris2017QuantumProcessor}. Such grid architectures are already arranged in node columns and thus can be calibrated efficiently by following the steps of Fig. \ref{fig:parallelcalibration}, potentially without embedded detectors.

\end{document}